\documentclass[ALICE,manyauthors]{cernphprep}
\usepackage[comma,square,numbers,sort&compress]{natbib}
\usepackage{hyperref}
\usepackage{lineno}
\usepackage{xspace}
\usepackage[T1]{fontenc}
\usepackage{orcidlink} 

\begin{document}
%

\newcommand{\pp}           {pp\xspace}
\newcommand{\ppbar}        {\mbox{$\mathrm {p\overline{p}}$}\xspace}
\newcommand{\XeXe}         {\mbox{Xe--Xe}\xspace}
\newcommand{\PbPb}         {\mbox{Pb--Pb}\xspace}
\newcommand{\pA}           {\mbox{pA}\xspace}
\newcommand{\pPb}          {\mbox{p--Pb}\xspace}
\newcommand{\AuAu}         {\mbox{Au--Au}\xspace}
\newcommand{\dAu}          {\mbox{d--Au}\xspace}

\newcommand{\s}            {\ensuremath{\sqrt{s}}\xspace}
\newcommand{\snn}          {\ensuremath{\sqrt{s_{\mathrm{NN}}}}\xspace}
\newcommand{\pt}           {\ensuremath{p_{\rm T}}\xspace}
\newcommand{\meanpt}       {$\langle p_{\mathrm{T}}\rangle$\xspace}
\newcommand{\ycms}         {\ensuremath{y_{\rm CMS}}\xspace}
\newcommand{\ylab}         {\ensuremath{y_{\rm lab}}\xspace}
\newcommand{\etarange}[1]  {\mbox{$\left | \eta \right |~<~#1$}}
\newcommand{\yrange}[1]    {\mbox{$\left | y \right |~<~#1$}}
\newcommand{\dndy}         {\ensuremath{\mathrm{d}N_\mathrm{ch}/\mathrm{d}y}\xspace}
\newcommand{\dndeta}       {\ensuremath{\mathrm{d}N_\mathrm{ch}/\mathrm{d}\eta}\xspace}
\newcommand{\avdndeta}     {\ensuremath{\langle\dndeta\rangle}\xspace}
\newcommand{\dNdy}         {\ensuremath{\mathrm{d}N_\mathrm{ch}/\mathrm{d}y}\xspace}
\newcommand{\Npart}        {\ensuremath{N_\mathrm{part}}\xspace}
\newcommand{\Ncoll}        {\ensuremath{N_\mathrm{coll}}\xspace}
\newcommand{\dEdx}         {\ensuremath{\textrm{d}E/\textrm{d}x}\xspace}
\newcommand{\RpPb}         {\ensuremath{R_{\rm pPb}}\xspace}

\newcommand{\nineH}        {$\sqrt{s}~=~0.9$~Te\kern-.1emV\xspace}
\newcommand{\seven}        {$\sqrt{s}~=~7$~Te\kern-.1emV\xspace}
\newcommand{\twoH}         {$\sqrt{s}~=~0.2$~Te\kern-.1emV\xspace}
\newcommand{\twosevensix}  {$\sqrt{s}~=~2.76$~Te\kern-.1emV\xspace}
\newcommand{\five}         {$\sqrt{s}~=~5.02$~Te\kern-.1emV\xspace}
\newcommand{\twosevensixnn}{$\sqrt{s_{\mathrm{NN}}}~=~2.76$~Te\kern-.1emV\xspace}
\newcommand{\fivenn}       {$\sqrt{s_{\mathrm{NN}}}~=~5.02$~Te\kern-.1emV\xspace}
\newcommand{\LT}           {L{\'e}vy-Tsallis\xspace}
\newcommand{\GeVc}         {Ge\kern-.1emV/$c$\xspace}
\newcommand{\MeVc}         {Me\kern-.1emV/$c$\xspace}
\newcommand{\TeV}          {Te\kern-.1emV\xspace}
\newcommand{\GeV}          {Ge\kern-.1emV\xspace}
\newcommand{\MeV}          {Me\kern-.1emV\xspace}
\newcommand{\GeVmass}      {Ge\kern-.2emV/$c^2$\xspace}
\newcommand{\MeVmass}      {Me\kern-.2emV/$c^2$\xspace}
\newcommand{\lumi}         {\ensuremath{\mathcal{L}}\xspace}

\newcommand{\ITS}          {\rm{ITS}\xspace}
\newcommand{\TOF}          {\rm{TOF}\xspace}
\newcommand{\ZDC}          {\rm{ZDC}\xspace}
\newcommand{\ZDCs}         {\rm{ZDCs}\xspace}
\newcommand{\ZNA}          {\rm{ZNA}\xspace}
\newcommand{\ZNC}          {\rm{ZNC}\xspace}
\newcommand{\SPD}          {\rm{SPD}\xspace}
\newcommand{\SDD}          {\rm{SDD}\xspace}
\newcommand{\SSD}          {\rm{SSD}\xspace}
\newcommand{\TPC}          {\rm{TPC}\xspace}
\newcommand{\TRD}          {\rm{TRD}\xspace}
\newcommand{\VZERO}        {\rm{V0}\xspace}
\newcommand{\VZEROA}       {\rm{V0A}\xspace}
\newcommand{\VZEROC}       {\rm{V0C}\xspace}
\newcommand{\Vdecay} 	   {\ensuremath{V^{0}}\xspace}

\newcommand{\ee}           {\ensuremath{e^{+}e^{-}}} 
\newcommand{\pip}          {\ensuremath{\pi^{+}}\xspace}
\newcommand{\pim}          {\ensuremath{\pi^{-}}\xspace}
\newcommand{\kap}          {\ensuremath{\rm{K}^{+}}\xspace}
\newcommand{\kam}          {\ensuremath{\rm{K}^{-}}\xspace}
\newcommand{\pbar}         {\ensuremath{\rm\overline{p}}\xspace}
\newcommand{\kzero}        {\ensuremath{{\rm K}^{0}_{\rm{S}}}\xspace}
\newcommand{\lmb}          {\ensuremath{\Lambda}\xspace}
\newcommand{\almb}         {\ensuremath{\overline{\Lambda}}\xspace}
\newcommand{\Om}           {\ensuremath{\Omega^-}\xspace}
\newcommand{\Mo}           {\ensuremath{\overline{\Omega}^+}\xspace}
\newcommand{\X}            {\ensuremath{\Xi^-}\xspace}
\newcommand{\Ix}           {\ensuremath{\overline{\Xi}^+}\xspace}
\newcommand{\Xis}          {\ensuremath{\Xi^{\pm}}\xspace}
\newcommand{\Oms}          {\ensuremath{\Omega^{\pm}}\xspace}
\newcommand{\degree}       {\ensuremath{^{\rm o}}\xspace}

\begin{titlepage}
\PHyear{2025}       
\PHnumber{084}      
\PHdate{02 April}   


\title{Long-range transverse momentum correlations and radial flow in Pb$-$Pb collisions at the LHC with ALICE}
\ShortTitle{Long-range transverse momentum correlations and radial flow}   

\Collaboration{ALICE Collaboration\thanks{See Appendix~\ref{app:collab} for the list of collaboration members}}
\ShortAuthor{ALICE Collaboration} 

\begin{abstract}
This Letter presents measurements of long-range transverse-momentum correlations using a new observable, $v_{0}(p_\mathrm{T})$, serving as a probe of event-by-event radial-flow fluctuations, the underlying radial expansion, and the medium’s properties in heavy-ion collisions. Results are reported for inclusive charged particles, pions, kaons, and protons across various centrality intervals in Pb$-$Pb collisions at $\sqrt{s_\mathrm{NN}} = 5.02$ TeV, recorded by the ALICE detector. A pseudorapidity-gap technique, similar to that used in anisotropic-flow studies, is employed to suppress short-range correlations. At low $p_\mathrm{T}$, a characteristic mass ordering consistent with hydrodynamic collective flow is observed. At higher $p_\mathrm{T}$ ($> 3$~GeV/$c$), protons exhibit larger $v_{0}(p_\mathrm{T})$ than pions and kaons, in agreement with expectations from quark-recombination models. Comparisons to viscous hydrodynamic calculations with varying bulk viscosity and equation of state demonstrate the sensitivity of the $v_{0}(p_\mathrm{T})$ observable to these key medium properties. The findings establish $v_{0}(p_\mathrm{T})$ as a valuable addition to the set of observables used in Bayesian analyses for extracting the transport properties and constraining the equation of state of strongly interacting matter, while also helping to systematically explore its sensitivity and impact within such global studies.

\end{abstract}
\end{titlepage}

\setcounter{page}{2} 



The study of collective behavior in high-energy heavy-ion collisions provides insights into the properties of the quark--gluon plasma (QGP), the deconfined state of quarks and gluons predicted by Quantum Chromodynamics (QCD)~\cite{Shuryak:1978ij, Shuryak:1980tp, Bass:1998vz, Cleymans:1985wb}. A key experimental signature of collectivity is the emergence of long-range correlations in azimuthal angle and pseudorapidity among produced particles, reflecting the response of the medium to the initial-state conditions~\cite{Ollitrault:1992bk, Reisdorf:1997fx, Voloshin:2008dg, Armesto:2004pt, PHOBOS:2008cdq, STAR:2009ngv, CMS:2011cqy, ALICE:2011svq}. The long-range azimuthal correlations arise from the initial spatial anisotropies, translated into momentum anisotropy by pressure gradients during the hydrodynamic expansion of the QCD medium~\cite{Ollitrault:1992bk, Voloshin:2008dg, Ollitrault:2023wjk}. Experiments at the Relativistic Heavy Ion Collider (RHIC) and the Large Hadron Collider (LHC) have measured these correlations across different systems and energies~\cite{STAR:2000ekf, PHENIX:2004vcz, ALICE:2010suc, ALICE:2011ab, ATLAS:2012at, CMS:2013wjq, ALICE:2016ccg, PHENIX:2018lia}. Hydrodynamic models, which assume local thermal equilibrium and use a lattice QCD (LQCD)–based equation of state (EOS), reproduce these measurements and help constrain the transport coefficients of the system produced in heavy-ion collisions~\cite{Heinz:2013th, Gale:2013da, Moreland:2018gsh, Gardim:2019brr, HotQCD:2014kol}.

The azimuthal correlations are quantified by the $v_n$ coefficients, obtained from the Fourier-series decomposition of the azimuthal-angle distribution in momentum space of final-state particles, with respect to the reaction plane spanned by the beam axis and the impact parameter~\cite{Voloshin:1994mz, Poskanzer:1998yz}. These coefficients quantify anisotropies in particle momentum distributions, which can arise from both collective effects, such as anisotropic flow~\cite{Teaney:2000cw, Qin:2010pf, Shen:2011eg, ALICE:2016ccg, ALICE:2022zks}, and non-equilibrium phenomena like jets~\cite{Gyulassy:1990ye, ATLAS:2013ssy, ALICE:2015efi, CMS:2017xgk, ALICE:2018rtz}. Alongside anisotropic expansion, the system also undergoes isotropic expansion, known as radial flow, which modifies the transverse-momentum ($p_\mathrm{T}$) distribution of the produced particles. Radial flow is often inferred from the slope of the $p_\mathrm{T}$ distribution, which is a convolution of thermal radiation and collective expansion of the medium. Commonly, the radial-flow parameter is extracted by fitting the experimental $p_\mathrm{T}$ distribution to Boltzmann-Gibbs blast-wave function, a simplified hydrodynamic model accounting for collective expansion~\cite{Schnedermann:1993ws, STAR:2009sxc, STAR:2017sal, ALICE:2019hno}. However, the $\langle\beta_\mathrm{T}\rangle$ parameter obtained from this approach is $p_\mathrm{T}$-integrated and thus does not provide direct access to $p_\mathrm{T}$-differential features such as the mass-ordering of $v_n(p_\mathrm{T})$ ($n \geq 2$) at low $p_\mathrm{T}$ and baryon-meson splitting observed at intermediate $p_\mathrm{T}$. Short-range correlations in pseudorapidity ($\eta$), such as those from resonance decays or near-side jets, are typically not suppressed in radial-flow measurements using inclusive $p_\mathrm{T}$ spectrum analysis. These correlations, which are unrelated to hydrodynamic collectivity, are referred to as nonflow. The recently introduced observable $v_{0}(p_\mathrm{T})$ reduces these effects, allowing radial flow to be studied in a manner similar to anisotropic flow~\cite{Schenke:2020uqq, Parida:2024ckk, ATLAS:2025ztg}. It is defined as the normalized covariance between event-by-event multiplicity and mean $p_\mathrm{T}$ of the event, evaluated using a pseudorapidity gap ($\Delta\eta$) to suppress nonflow contributions while preserving long-range $p_\mathrm{T}$ correlations.  Following Ref.~\cite{Parida:2024ckk}, $v_{0}(p_\mathrm{T})$ is defined as
\begin{equation}
\label{eq:formula1}
v_{0}(p_\mathrm{T}) = \frac{\langle f_{A}(p_\mathrm{T})[p_\mathrm{T}]_{B}\rangle - \langle f_{A}(p_\mathrm{T})\rangle\langle [p_\mathrm{T}]_{B}\rangle}{\langle f_{A}(p_\mathrm{T})\rangle\sigma_{[p_\mathrm{T}]}},
\end{equation}
where 
\begin{equation}
\label{eq:formula2}
\sigma_{[p_\mathrm{T}]}=\sqrt{\langle [p_\mathrm{T}]_{A}[p_\mathrm{T}]_{B}\rangle - \langle [p_\mathrm{T}]_{A}\rangle\langle [p_\mathrm{T}]_{B}\rangle}.
\end{equation}
In the above equations, $A$ and $B$ represent two distinct $\eta$ windows with a separation of $\Delta\eta$. The function $f_{A}(p_\mathrm{T})$ represents the fraction of particles in each $p_\mathrm{T}$ bin relative to the total number of particles in an event within the $\eta$ window $A$. The mean $p_\mathrm{T}$ calculated within one event in $\eta$ windows $A$ and $B$, are denoted as $[p_\mathrm{T}]_{A}$ and $[p_\mathrm{T}]_{B}$, respectively. The brackets $\langle ..\rangle$ indicate an average taken over all events. In hydrodynamics, long-range $p_\mathrm{T}$ correlations and azimuthal correlations share a common origin, both arising from the pressure gradients. While the initial spatial geometry drives azimuthal anisotropies in particle distributions, the fireball's initial temperature and size influence the shape of the $p_\mathrm{T}$ spectra. The observable $v_0(p_\mathrm{T})$ quantifies modifications to the spectra due to variations in radial flow, and since it reflects the isotropic expansion of the system, it is sensitive to the bulk viscosity of the medium. Moreover, the relation between $\langle\beta_\mathrm{T}\rangle$ and $v_{0}(p_\mathrm{T})$ can be illustrated using the framework of the blast-wave model~\cite{Saha:2025nyu}, as detailed in the supplementary material~\ref{sec:blast}. 

Hydrodynamic simulations have shown that $v_{0}(p_\mathrm{T})$ exhibits the following features: (a) $v_{0}(p_\mathrm{T})$ changes sign with $p_\mathrm{T}$\textemdash{}negative at low $p_\mathrm{T}$ and positive at high $p_\mathrm{T}$~\cite{Parida:2024ckk, Schenke:2020uqq}. This reflects the correlation between event-by-event mean-$p_\mathrm{T}$ fluctuations and the spectral shape. An upward fluctuation in mean $p_\mathrm{T}$ ($[p_\mathrm{T}]>\langle[p_\mathrm{T}]\rangle$) increases the fraction of high-$p_\mathrm{T}$ particles and decreases that of low-$p_\mathrm{T}$ particles, while a downward fluctuation ($[p_\mathrm{T}]<\langle[p_\mathrm{T}]\rangle$) does the opposite. Similar behavior appears in a simple toy model~\cite{Schenke:2020uqq} ($p_\mathrm{T}$ spectra of exponential form, $\mathrm{d}N/\mathrm{d}p_\mathrm{T} = (2p_\mathrm{T}N)\frac{\mathrm{exp}(-2p_\mathrm{T}/[p_\mathrm{T}])}{[p_\mathrm{T}]^{2}}$ with fluctuating parameters, $N$ and $[p_\mathrm{T}]$), where $v_{0}(p_\mathrm{T})$ simplifies to $v_{0}(p_\mathrm{T})\approx2\frac{\sigma_{[p_\mathrm{T}]}}{\langle[p_\mathrm{T}]\rangle}\left(\frac{p_\mathrm{T}}{\langle[p_\mathrm{T}]\rangle}-1\right)$, showing that the sign change naturally arises from mean-$p_\mathrm{T}$ fluctuations ($\sigma_{[p_\mathrm{T}]}$), whose physical origin can be different in each model, depending on its assumptions and mechanisms; (b) $v_{0}(p_\mathrm{T})$ for identified particles shows species dependence and mass-ordering~\cite{Parida:2024ckk, Schenke:2020uqq}, similar to that observed for $v_{2}(p_\mathrm{T})$~\cite{STAR:2001ksn, PHENIX:2003qra, ALICE:2014wao, ALICE:2022zks} and higher harmonics of anisotropic flow~\cite{PHENIX:2014uik, ALICE:2016cti, ALICE:2022zks}. This indicates $v_{0}(p_\mathrm{T})$, a measure of radial flow and its fluctuations, captures hydrodynamic effects~\cite{Huovinen:2001cy, Teaney:2000cw, Ollitrault:2007du, Borghini:2005kd, Parida:2024ckk}; (c) $v_{0}(p_\mathrm{T})$ exhibits a centrality dependence, scaling approximately as $(\sqrt{\mathrm{d}N_\text{ch}/\mathrm{d}\eta})^{-1}$~\cite{Schenke:2020uqq}. A scaled observable, $v_{0}(p_\mathrm{T})/v_{0}$\textemdash{}where $v_{0}$ is calculated as $\sigma_{[p_\mathrm{T}]}/\langle [p_\mathrm{T}]\rangle$\textemdash{}is found to be independent of system size at a given collision energy~\cite{Parida:2024ckk}. Theoretically, this scaling reduces the dependence on the absolute size of fluctuations~\cite{Parida:2024ckk}, similar to the scaled anisotropic flow, $v_{n}(p_\mathrm{T})/v_{n}$~\cite{ATLAS:2018ezv}; (d) $v_0(p_\mathrm{T})$ is sensitive to the QCD medium’s bulk viscosity and the equation of state~\cite{Parida:2024ckk, Gong:2024lhq}. However, since it is influenced by the radial expansion of the system, it remains mostly unaffected by variations in shear viscosity. Furthermore, $v_0(p_\mathrm{T})$ is closely related to $\sigma_{[p_\mathrm{T}]}$, as shown in Eq.~\ref{eq:formula1}. Unlike previous measurements of $\sigma_{[p_\mathrm{T}]}$ across various systems and collision energies~\cite{NA49:1999inh, PHENIX:2002aqz, PHENIX:2003ccl, STAR:2003cbv, NA49:2003hxt, CERES:2003sap, STAR:2005vxr, STAR:2005xhg, NA49:2008fag, STAR:2013sov, ALICE:2014gvd, STAR:2019dow, ATLAS:2022dov, ALICE:2021gxt, ALICE:2023tej, ATLAS:2024jvf, ALICE:2024apz}, this study employs a $p_\mathrm{T}$-differential approach with a $\Delta\eta$ gap to suppress nonflow effects. While the $\Delta\eta$-gap method effectively reduces short-range, near-side jet-like correlations, it does not eliminate long-range away-side correlations~\cite{ALICE:2023lyr}, which may impact $v_0(p_\mathrm{T})$ measurements. 

The first experimental measurement of $v_{0}(p_\mathrm{T})$ for inclusive charged particles ($\mathrm{h}^{\pm}$), pions ($\pi^{\pm}$), kaons ($\mathrm{K}^{\pm}$), and protons ($\mathrm{p}(\bar{\mathrm{p}}$)) across various centrality intervals in Pb$-$Pb collisions is reported in the Letter. The results are obtained from a data sample of Pb--Pb collisions at $\sqrt{s_\mathrm{NN}} = 5.02$ TeV, collected by ALICE at the LHC in 2018. Details of the ALICE detector and its performance can be found in Refs.~\cite{ALICE:2008ngc, ALICE:2014sbx}. Minimum-bias events are triggered via coincidence signals in the V0 detector~\cite{ALICE:2013axi, ALICE:2004ftm}, which consists of two scintillator arrays, V0A and V0C, covering the ranges $2.8 < \eta < 5.1$ and $-3.7 < \eta < -1.7$, respectively. Only events with a reconstructed primary vertex within $\pm 10$ cm along the beam direction from the nominal interaction point are selected, and events with more than one reconstructed primary interaction vertex (pileup events~\cite{Arslandok:2022dyb}) are excluded. Approximately 80 million minimum-bias collisions are selected for analysis, and categorized into centrality intervals based on the amplitude distribution measured in the V0 detector~\cite{ALICE:2013hur}.

The charged particle tracks are reconstructed in the ALICE central barrel within $|\eta| < 0.8$, $0.2 < p_\mathrm{T} < 10.0$~GeV/$c$, and in the full azimuth, using the Inner Tracking System (ITS)~\cite{ALICE:2010tia}, the Time Projection Chamber (TPC)~\cite{Alme:2010ke}, and the Time-Of-Flight (TOF)~\cite{ALICE:2000xcm} detector. Particles with at least one space point in the two innermost layers of the ITS and a minimum of 70 out of 159 space points in the TPC are selected. The chi-square ($\chi^2$) per space point in the TPC and the ITS resulting from the track fit is required to be below 2.5 and 36, respectively. Tracks with $p_\mathrm{T} < 0.2$~GeV/$\it{c}$ are rejected due to low tracking efficiency. For protons, an additional minimum-$p_\mathrm{T}$ threshold of 0.4~GeV/$\it{c}$ is applied to reduce contributions from secondary protons generated by interactions of charged particles with the detector material. To further minimize contamination from secondary particles, a criterion on the maximum distance of closest approach (DCA) of the track to the collision point of less than 2 cm in longitudinal direction and a $p_\mathrm{T}$-dependent selection in the transverse direction (less than $0.0105+0.035/p_\mathrm{T}^{1.1}$ in cm based on $p_\mathrm{T}$ of the track expressed in units of GeV/$c$) is applied. The width of the pseudorapidity gap is set to $\Delta\eta = 0.4$ to suppress short-range nonflow contributions (studies with different $\Delta\eta$ are presented in the supplementary material~\ref{sec:etadep}).

The identification of $\pi^{\pm}$, $\mathrm{K}^{\pm}$, and $\mathrm{p}(\bar{\mathrm{p}})$ is based on the specific energy loss ($\mathrm{d}E/\mathrm{d}x$) measured by the TPC and the time of flight from
the TOF, using the normalized deviations of the measured signals from the expected values for each species ($\sigma_\mathrm{TPC}$ and $\sigma_\mathrm{TOF}$, respectively). A Bayesian approach combines these signals with species-specific priors, yielding probabilities used for particle selection~\cite{ALICE:2016zzl}. Minimum-probability thresholds ($P_\mathrm{th}^\mathrm{min}$) of 0.95 for $\pi^{\pm}$, and 0.9 for $\mathrm{K}^{\pm}$ and $\mathrm{p}(\bar{\mathrm{p}})$ are applied. Additionally, particles are required to satisfy $|\mathrm{n\sigma}_\mathrm{TPC}| < 3$ and $|\mathrm{n\sigma}_\mathrm{TOF}| < 3$ over the entire $p_\mathrm{T}$ range. This method ensures high purity while minimizing misidentification effects. The $p_\mathrm{T}$-dependent purity estimated using Monte Carlo (MC) simulations is higher than 98\% (97\%) for $\pi^{\pm}$ ($\mathrm{p}(\bar{\mathrm{p}})$) in the $p_\mathrm{T}$-range $0.2 < p_\mathrm{T} < 6.0$~GeV/$\it{c}$ ($0.4 < p_\mathrm{T} < 6.0$~GeV/$\it{c}$). For $\mathrm{K}^{\pm}$, the purity is higher than 95\% in $0.2 < p_\mathrm{T} < 4.0$~GeV/$\it{c}$, and is nearly 90\% in $4.0 < p_\mathrm{T} < 6.0$~GeV/$\it{c}$.

The observable $v_0(p_\mathrm{T})$ is unaffected by tracking and particle-identification (PID) inefficiencies, and thus no efficiency corrections are applied. This robustness is validated through an MC closure test using events generated with the Heavy-Ion Jet Interaction Generator (HIJING)~\cite{Wang:1991hta, ToporPop:2004lve}, transported via GEANT3~\cite{Brun:1994aa}, and reconstructed with the same procedure as experimental data (see supplementary material~\ref{sec:mc_closure}). Statistical uncertainties are determined using the bootstrap sampling method~\cite{beran1991asympotic}, while systematic uncertainties on $v_{0}(p_\mathrm{T})$ are evaluated by varying event selection, track selection, and PID criteria. The uncertainties are computed as a function of $p_\mathrm{T}$, for each centrality interval. Event-selection uncertainties are assessed by modifying the primary vertex position acceptance and relaxing pileup-rejection criteria. Uncertainties associated with centrality estimation are addressed by redefining centrality intervals based on the multiplicity distribution measured at midrapidity~\cite{ALICE:centrality1}. Track-selection uncertainties are determined by varying the DCA criteria in both longitudinal and transverse directions, the number of reconstructed space points in the TPC, and track fit quality requirements. Variations in $\Delta\eta$ gap are also explored to assess their impact on nonflow suppression. PID-related uncertainties for $\pi^{\pm}$, $\mathrm{K}^{\pm}$, and $\mathrm{p}(\bar{\mathrm{p}})$ are estimated by varying the default $P_\mathrm{th}^{\text{min}}$. All systematic uncertainty sources are treated as uncorrelated, and the total systematic uncertainty is obtained by summing their contributions in quadrature. Percentage contributions of various sources to the total systematic uncertainty for a representative centrality interval are provided in the supplementary material~\ref{sec:app_sys}.


\begin{figure}[t]
  \begin{center}
    \includegraphics[width=0.78\linewidth]{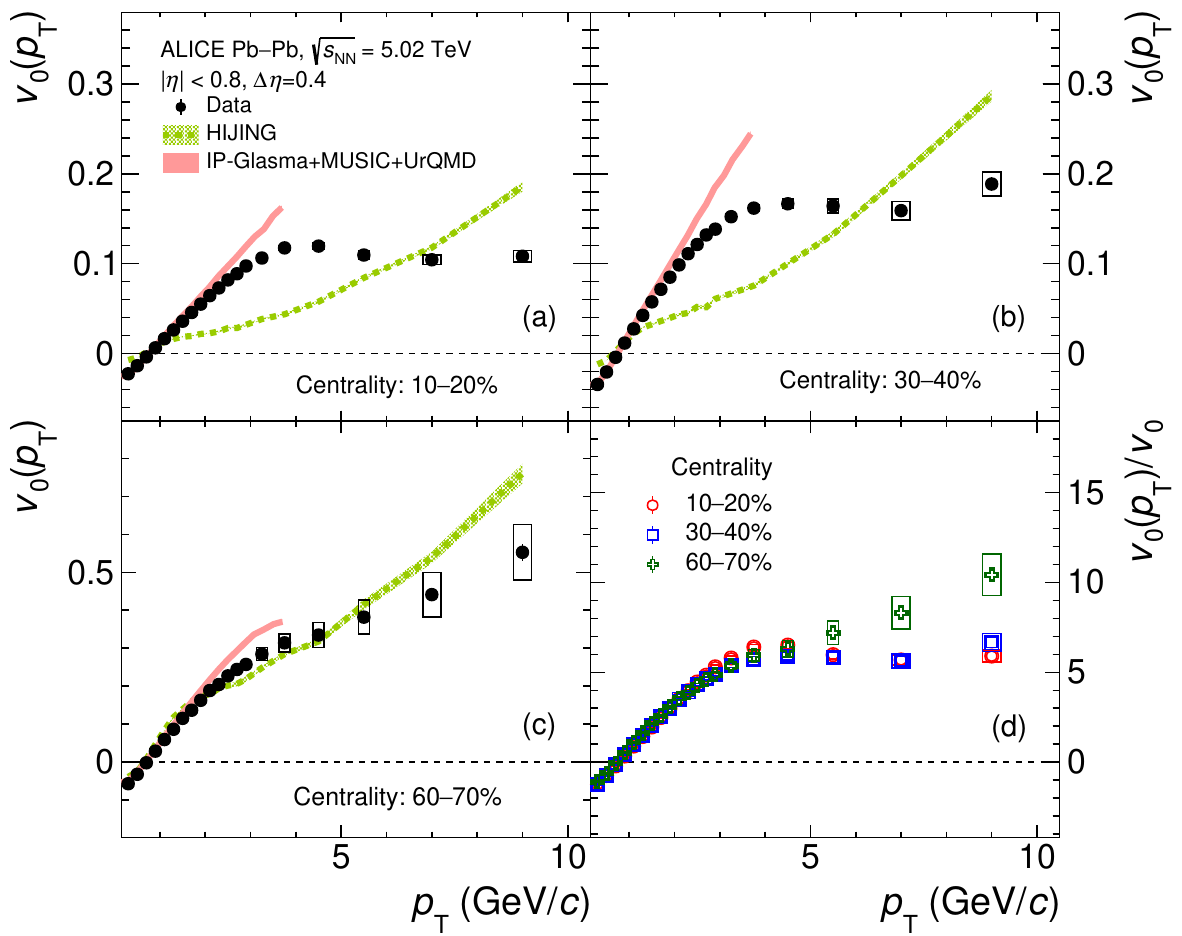}
  \end{center}
  \caption{$v_{0}(p_\mathrm{T})$ of inclusive charged particles shown as a function of $p_\mathrm{T}$ in Pb$-$Pb collisions at $\sqrt{s_\mathrm{NN}} = 5.02$ TeV for centrality intervals 10$-$20\% (a), 30$-$40\% (b), and 60$-$70\% (c). The measurements are compared to expectations from HIJING~\cite{Wang:1991hta} and IP-Glasma+MUSIC+UrQMD~\cite{Mantysaari:2024uwn} models. $v_{0}(p_\mathrm{T})/v_{0}$ of inclusive charged particles shown as a function of $p_\mathrm{T}$ for the centrality intervals (d). The statistical (systematic) uncertainties are represented by vertical bars (boxes).}
  \label{fig:v0pT_had}
\end{figure}

\begin{figure}[t]
  \begin{center}
    \includegraphics[width=\linewidth]{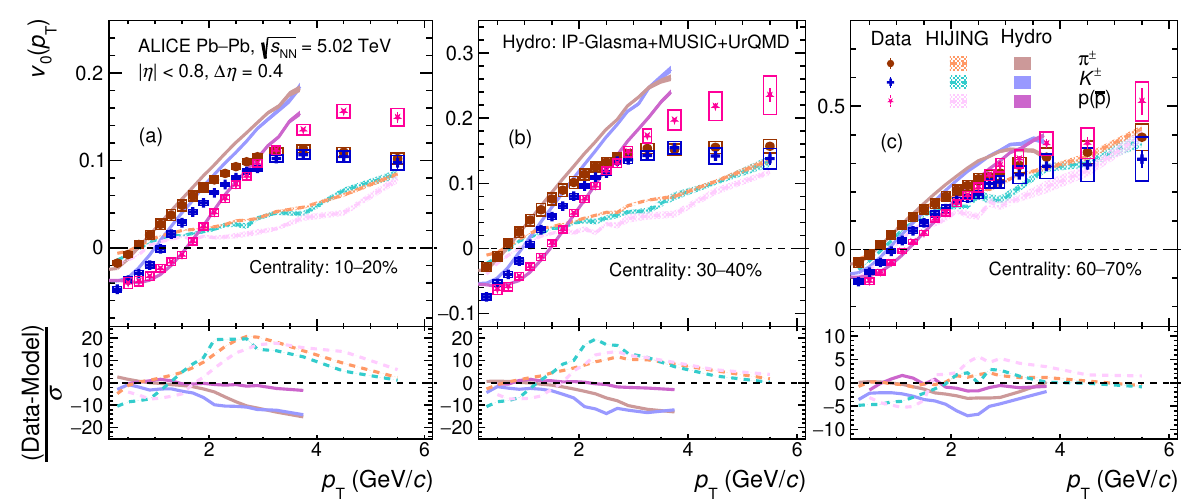}
  \end{center}
  \caption{$v_{0}(p_\mathrm{T})$ of pions ($\pi^{\pm}$), kaons ($\mathrm{K}^{\pm}$), and protons ($\mathrm{p}(\bar{\mathrm{p}})$) shown as a function of $p_\mathrm{T}$ in Pb$-$Pb collisions at $\sqrt{s_\mathrm{NN}} = 5.02$ TeV for centrality intervals 10$-$20\% (a), 30$-$40\% (b), and 60$-$70\% (c). The measurements are compared to results from HIJING~\cite{Wang:1991hta} and IP-Glasma+MUSIC+UrQMD~\cite{Mantysaari:2024uwn} models. The statistical (systematic) uncertainties are represented by vertical bars (boxes). The bottom panels show the (Data-Model)/$\sigma$, representing the deviation between the experimental data and model predictions, normalized by the uncertainty.}
  \label{fig:v0pT_piKprot}
\end{figure}

The evolution of $v_{0}(p_\mathrm{T})$ for inclusive charged particles is presented in panels (a), (b) and (c) of Fig.~\ref{fig:v0pT_had} for three centrality intervals: central (10$-$20\%), semicentral (30$-$40\%), and peripheral (60$-$70\%). The $v_{0}(p_\mathrm{T})$ is negative at low $p_\mathrm{T} (< $0.8~GeV/$\it{c}$) across all centralities. This is consistent with the anti-correlation between event-by-event mean-$p_\mathrm{T}$ fluctuations and particle production at different $p_\mathrm{T}$ as discussed above~\cite{Schenke:2020uqq, Parida:2024ckk}. For $p_\mathrm{T} < 4.0$~GeV/$c$, $v_{0}(p_\mathrm{T})$ exhibits an approximately linear increase with $p_\mathrm{T}$, with a slope that grows from central to peripheral collisions. The linear $p_\mathrm{T}$ dependence is similar to the predictions from the simple toy model~\cite{Schenke:2020uqq}. However, for $p_\mathrm{T}>4.0$~GeV/$\it{c}$, the data deviates from this linearly-increasing trend, with a clear decrease in the slope of $v_{0}(p_\mathrm{T})$ in central and semicentral collisions. In peripheral collisions, this change is much smaller, which may reflect differences in the relative contribution of hard and soft processes at high $p_\mathrm{T}$ compared to central and semicentral collisions. 

The data are compared to hydrodynamic model calculations from the IP-Glasma+MUSIC+UrQMD framework, which successfully describe the ALICE measurements of charged hadron and identified particle yields, mean $p_\mathrm{T}$, and anisotropic-flow coefficients~\cite{Mantysaari:2024uwn}. This model employs IP-Glasma initial conditions~\cite{Schenke:2012wb}, MUSIC hydrodynamic evolution~\cite{Schenke:2010nt}, and a hadronic cascade (UrQMD)~\cite{Bass:1998ca, Bleicher:1999xi}, incorporating a temperature-dependent specific shear ($\eta/s$) and bulk ($\zeta/s$) viscosity for the QGP~\cite{Mantysaari:2024uwn}. It includes an equation of state based on LQCD calculations~\cite{HotQCD:2014kol}. Across all centralities, including peripheral collisions, the model describes the data well up to $p_\mathrm{T} \approx 2.0$~GeV/$c$, beyond which deviations appear, similar to those observed in $v_2$ and $v_3$~\cite{ALICE:2016ccg, ALICE:2018rtz}. These deviations may indicate limitations of the hydrodynamic framework in capturing the transition from a strongly coupled medium to a more kinetic regime where hard processes and jet-medium interactions become relevant~\cite{Kolb:2003dz, Noronha-Hostler:2016eow}. The HIJING model~\cite{Wang:1991hta, ToporPop:2004lve} includes mini-jet production, resonance decays, and nuclear effects but lacks collective flow. As a result, mean-$p_\mathrm{T}$ fluctuations arising from such sources can lead to a sign change of $v_{0}(p_\mathrm{T})$ in this model. HIJING qualitatively describes high-$p_\mathrm{T}$ particle production from initial hard scatterings but underpredicts low-$p_\mathrm{T}$ yields and mean $p_\mathrm{T}$ due to the absence of medium-induced collective dynamics. Consequently, it fails to describe $v_{0}(p_\mathrm{T})$ in central and semicentral collisions, where medium-induced collective expansion is expected to dominate. However, in peripheral collisions, where the system size and energy density are lower, HIJING qualitatively captures the trend and magnitude of the data up to high $p_\mathrm{T}$, suggesting an increased role of hard scatterings and jet production.

The panel (d) of Fig.~\ref{fig:v0pT_had} presents the $p_\mathrm{T}$ dependence of the scaled observable $v_{0}(p_\mathrm{T})/v_{0}$ for the different centrality intervals. In central and semicentral collisions, the scaling behavior remains consistent with hydrodynamic expectations~\cite{Parida:2024ckk}. However, in peripheral collisions, minor deviations from this scaling appear for $p_\mathrm{T} > 5$~GeV/$c$, indicating the increasing influence of effects beyond collective flow, such as back-to-back jets and mini-jet fragmentation. The agreement between the measured $v_{0}(p_\mathrm{T})$ values and HIJING model predictions for peripheral collisions further supports that these non-collective contributions play a dominant role in shaping the observed deviations. A data-driven study to estimate such effects is performed by measuring $v_{0}(p_\mathrm{T})$ in azimuthal angle ranges, as presented in the supplementary material~\ref{sec:phidep}. 

Figure~\ref{fig:v0pT_piKprot} shows $v_{0}(p_\mathrm{T})$ as a function of $p_\mathrm{T}$ up to 6~GeV/$\it{c}$ for pions, kaons, and protons in three centrality intervals. The overall $p_\mathrm{T}$ dependence follows the trend observed for $\mathrm{h}^{\pm}$. A clear mass ordering is observed for $p_\mathrm{T} < 3$~GeV/$\it{c}$ across all centralities, consistent with expectations from the hydrodynamic model~\cite{Ollitrault:2007du, Borghini:2005kd, Parida:2024ckk}. For $p_\mathrm{T} > 3$~GeV/$\it{c}$, $v_{0}(p_\mathrm{T})$ for protons surpasses that of pions and kaons, with the latter two being consistent within uncertainties. This behavior is similar to the baryon-meson splitting observed for $v_n$~\cite{PHENIX:2003qra, ALICE:2014wao, ALICE:2022zks, PHENIX:2014uik, ALICE:2016cti}, suggesting quark recombination as the particle-production mechanism~\cite{Wan:2025rzg}. The separation between the results for protons and mesons ($\pi^{\pm}$ and $\mathrm{K}^{\pm}$) is significant in central and semicentral collisions, but becomes less pronounced in peripheral collisions. When $v_{0}(p_\mathrm{T})/n_q$ is plotted as a function of the transverse kinetic energy per constituent quark, $(m_\mathrm{T}-m_{0})/n_{q}$, the results for all species approximately collapse onto a common curve, demonstrating constituent quark number (NCQ) scaling~\ref{sec:NCQ}. This scaling behavior is consistent with the presence of partonic collectivity prior to hadronization, with hadrons forming via recombination of flowing quarks. The measurements are compared to theoretical predictions from IP-Glasma+MUSIC+UrQMD and HIJING. The hydrodynamic model captures the mass-dependent hierarchy of protons, kaons, and pions observed in the data across all centralities. The best agreement is found for protons, extending up to 3~GeV/$c$, while pions and kaons are described up to $\sim$2~GeV/$c$ and $\sim$1.5~GeV/$c$, with increasing deviations at higher $p_\mathrm{T}$. On the other hand, the HIJING model fails to reproduce the data in central and semicentral collisions. It neither captures the pronounced mass ordering characteristic of the hydrodynamic model, nor the observed separation between protons and mesons at higher $p_\mathrm{T}$ ($p_\mathrm{T} > 3$~GeV/$c$). In peripheral collisions, HIJING describes the overall trend for pions, but overestimates the values for kaons and protons, especially at low $p_\mathrm{T}$. 

\begin{figure}[th]
  \begin{center}
    \includegraphics[width=0.325\linewidth]{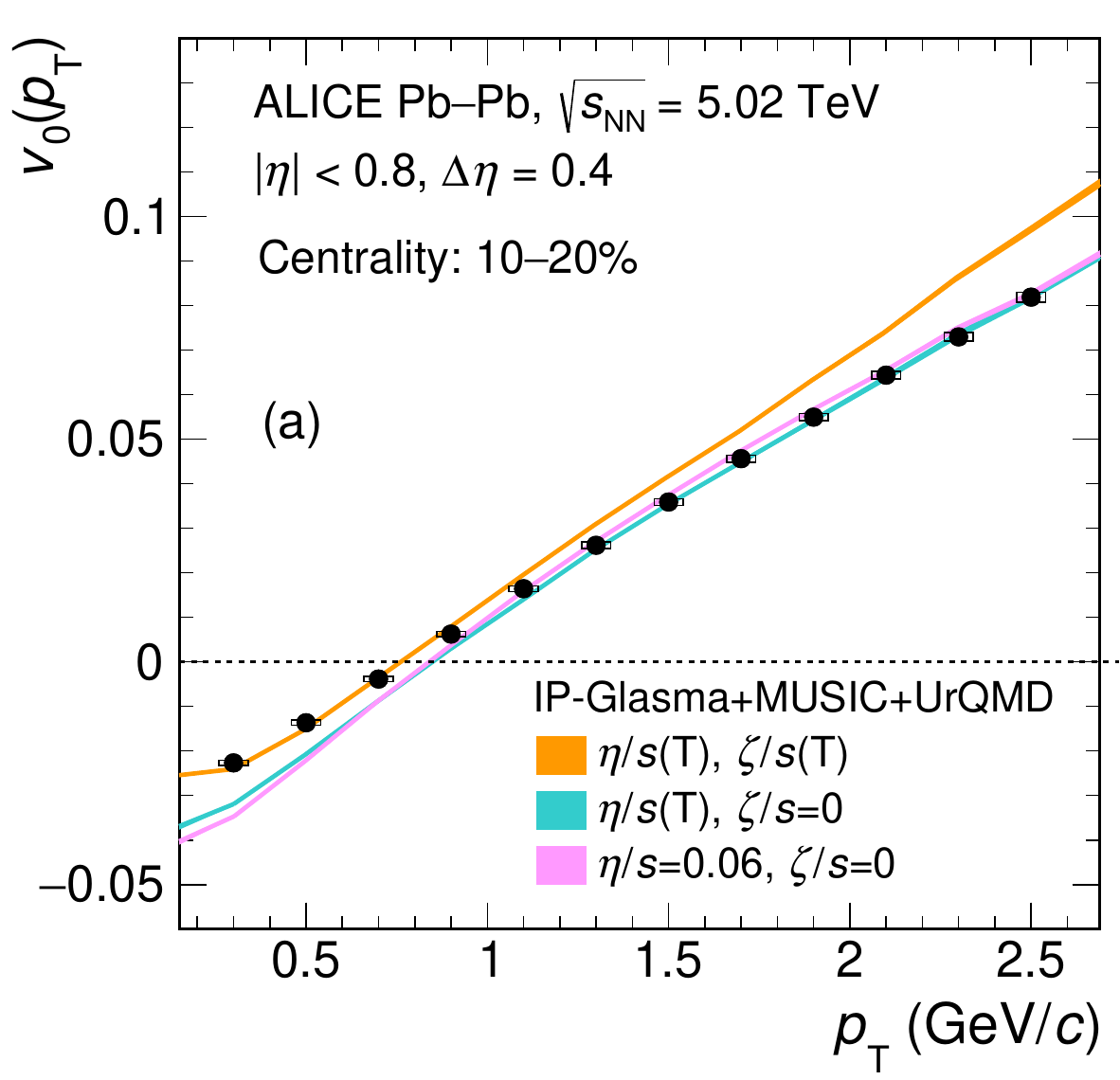}
    \includegraphics[width=0.325\linewidth]{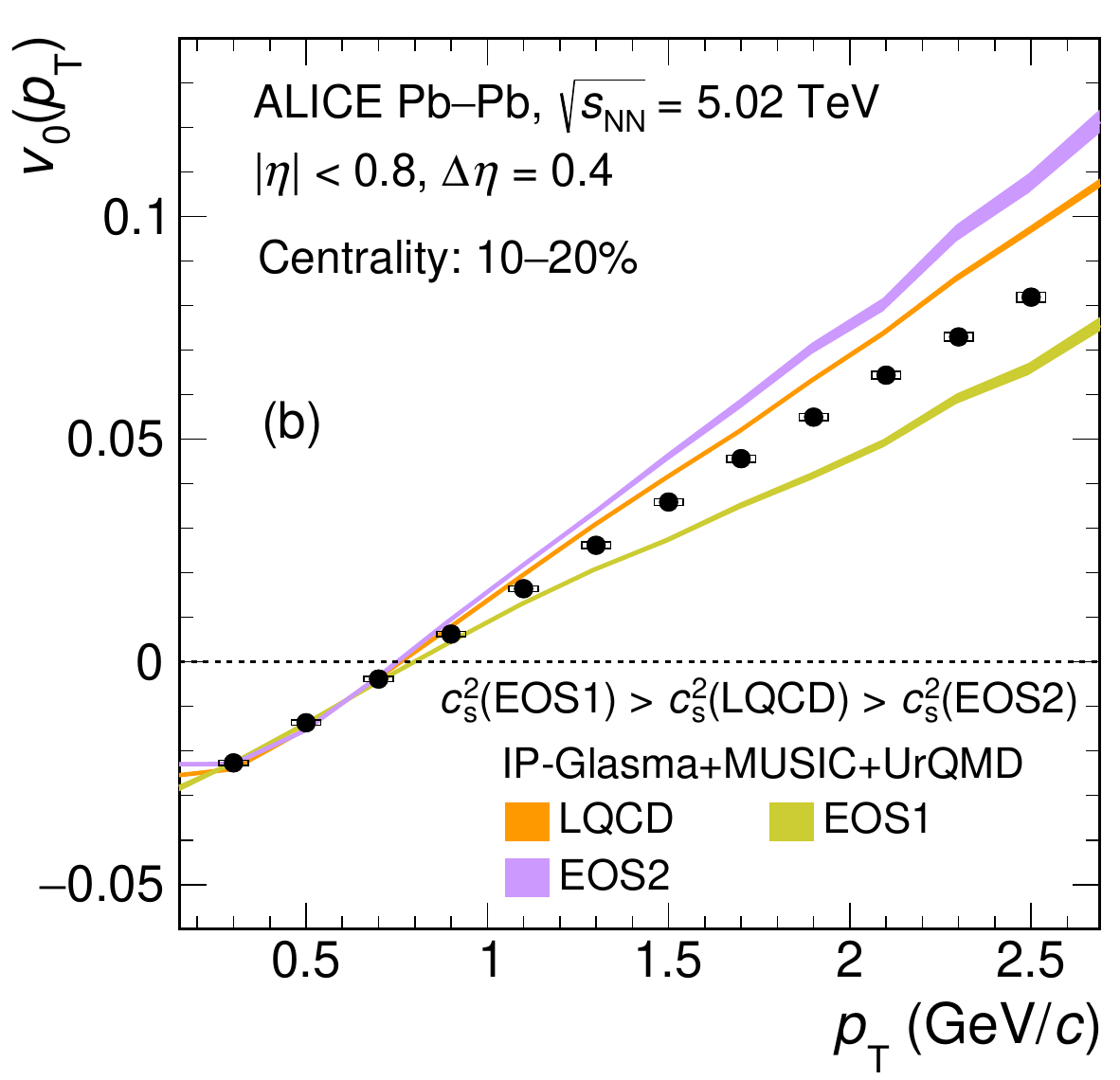}
    \includegraphics[width=0.325\linewidth]{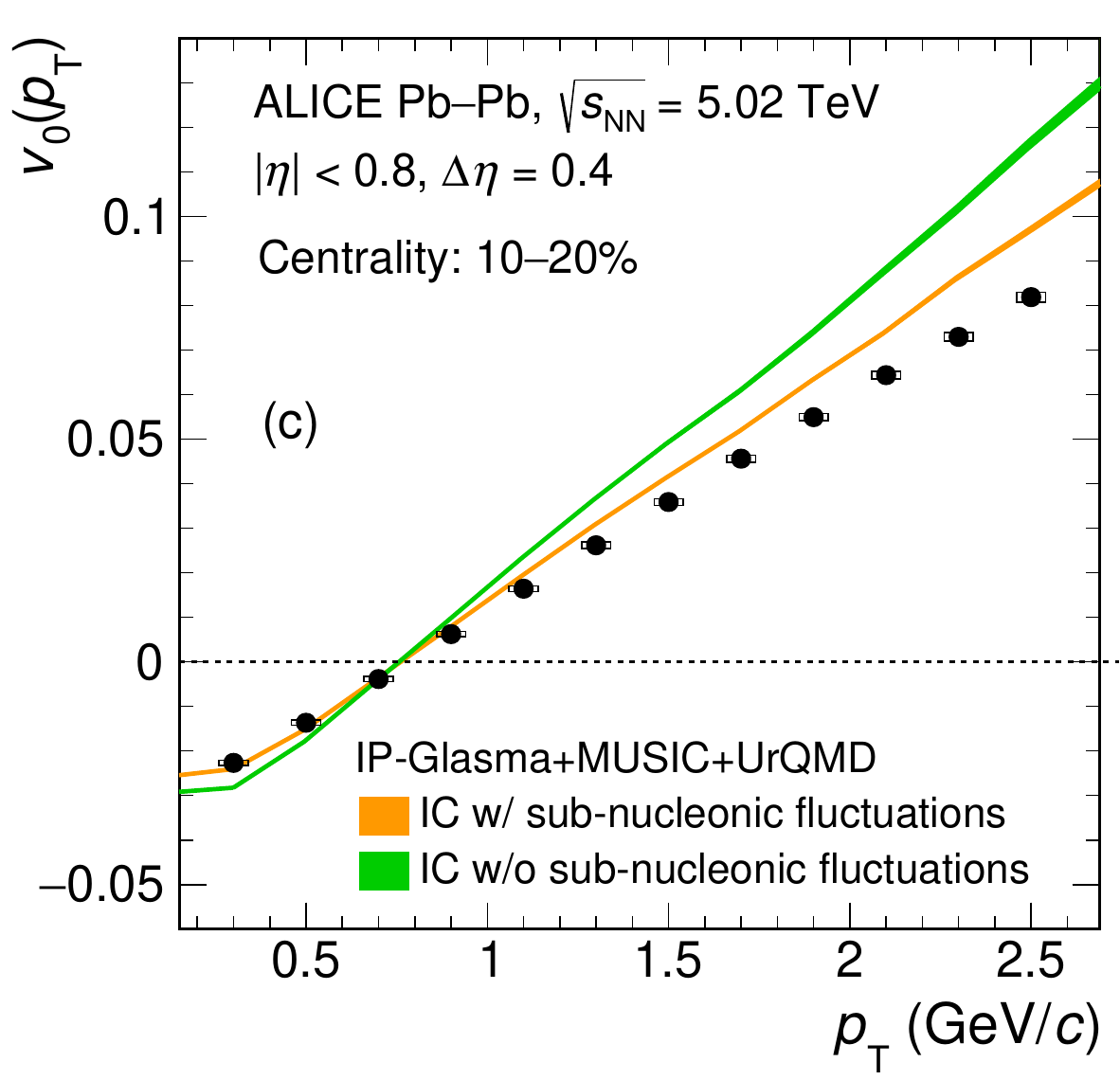}
  \end{center}
  \caption{Comparison of $v_{0}(p_\mathrm{T})$ for inclusive charged particles in 10$-$20\% central Pb$-$Pb collisions at $\sqrt{s_\mathrm{NN}} = 5.02$ TeV with IP-Glasma+MUSIC+UrQMD~\cite{Mantysaari:2024uwn} calculations: variations in transport coefficients, $\eta/s$ and $\zeta/s$ (a), different equations of state from Ref.~\cite{Gong:2024lhq} (b), and initial conditions (IC) with (w/) and without (w/o) sub-nucleonic fluctuations (c). The statistical (systematic) uncertainties are represented by vertical bars (boxes).}
  \label{fig:v0pT_etazeta_EOS}
\end{figure}
 
Figure~\ref{fig:v0pT_etazeta_EOS} shows the sensitivity of $v_{0}(p_\mathrm{T})$ to transport coefficients, EOS, and initial conditions for the centrality interval 10--20\%. The data shown are the same as those in the panel (a) of Fig.~\ref{fig:v0pT_had}, but zoomed into the low-$p_\mathrm{T}$ region. The panel (a) compares $v_{0}(p_\mathrm{T})$ for $\mathrm{h}^{\pm}$ with hydrodynamic predictions considering three scenarios: (i) both shear viscosity ($\eta/s$) and bulk viscosity ($\zeta/s$) are temperature dependent, (ii) $\eta/s$ is temperature dependent, while $\zeta/s = 0$, and (iii) $\eta/s$ = 0.06 and $\zeta/s = 0$. The model setup for (i) is exactly the same as in Figs.~\ref{fig:v0pT_had} and~\ref{fig:v0pT_piKprot}. The temperature dependence of $\eta/s$ and $\zeta/s$ over 150–400 MeV is detailed in Ref.~\cite{Bernhard:2016tnd, Bernhard:2019bmu, Mantysaari:2024uwn}. The chosen $\zeta/s$ was optimized to reproduce the centrality dependence of mean $p_\mathrm{T}$ of identified particles in Pb$-$Pb collisions at $\sqrt{s_\mathrm{NN}} = 5.02$ TeV~\cite{Mantysaari:2024uwn}. The predictions from scenarios (ii) and (iii), where $\zeta/s$ is fixed to zero but $\eta/s$ is either temperature-dependent or constant, are similar. In contrast, scenario (i), which includes a temperature-dependent $\zeta/s$, deviates from the other two. This suggests that $v_{0}(p_\mathrm{T})$ is primarily influenced by $\zeta/s$, unlike other observables such as $v_n$ and $p_\mathrm{T}$-spectra, which are sensitive to both. The sensitivity of $v_0(p_\mathrm{T})$ to $\zeta/s$ arises because $\zeta/s$ governs the system's resistance to isotropic expansion, thereby influencing the development of radial flow. Hydrodynamic calculations with temperature-dependent $\zeta/s$ describe the data better at low $p_\mathrm{T}$, while the higher-$p_\mathrm{T}$ region ($> 1.2$~GeV/$c$) may require further theoretical refinements.

The panel (b) of Fig.~\ref{fig:v0pT_etazeta_EOS} compares the same measurements with hydrodynamic predictions, using three EOS parametrizations, EOS1, EOS2, and the LQCD-based EOS (same orange curve in panel (a)). EOS1 and EOS2, derived for QCD matter at zero net-baryon density using a Gaussian Process Regression model constrained by LQCD calculations~\cite{Gong:2024lhq}, include transport coefficients ($\eta/s$ and $\zeta/s$) that satisfy causality conditions in relativistic viscous hydrodynamics. In the temperature range of 150$-$250 MeV, EOS1 exhibits a significantly larger squared speed of sound ($c_{s}^{2}$) compared to EOS2, while the LQCD-based EOS lies in between. A larger $c_{s}^{2}$ drives faster, more uniform expansion that enhances radial flow but reduces radial flow fluctuations and thus the $v_0(p_\mathrm{T})$ slope~\cite{Gong:2024lhq}. As a result, the slopes follow a reverse ordering relative to $c_{s}^{2}$, i.e., $\mathrm{slope}_\mathrm{EOS2}$ $>$ $\mathrm{slope}_\mathrm{LQCD}$ $>$ $\mathrm{slope}_\mathrm{EOS1}$. For $p_\mathrm{T} < 1$~GeV/$c$, $v_{0}(p_\mathrm{T})$ is unaffected by change in EOS, and model predictions agree well with data. Above 1~GeV/$c$, EOS2 overestimates the data, while EOS1 underestimates, highlighting the sensitivity of $v_{0}(p_\mathrm{T})$ to the underlying EOS. The LQCD-based EOS provides the best overall description of data. 

The panel (c) illustrates the sensitivity of $v_0(p_\mathrm{T})$ to initial energy density profile by comparing hydrodynamic simulations with and without sub-nucleonic fluctuations. The IP-Glasma model incorporates fluctuations at multiple length scales, arising from both the nuclear geometry and sub-nucleonic parton distributions~\cite{Mantysaari:2024uwn}. These fine spatial features alter the initial pressure gradients and hence influence the development of radial flow and its fluctuations. The steeper slope of $v_0(p_\mathrm{T})$ in the absence of sub-nucleonic fluctuations is consistent with the results obtained with a different initial-state model T\raisebox{-0.3ex}{R}ENTo~\cite{Moreland:2014oya} in Ref.~\cite{Parida:2024ckk}. The presence of sub-nucleonic fluctuations yields an improved representation of the data.

A more comprehensive understanding of transport coefficients and the EOS requires exploring multiple observables. While variations in the EOS, $\zeta/s$, and initial conditions can also affect other observables, such as $\mathrm{d}N/\mathrm{d}\eta$, $p_\mathrm{T}$ spectra, mean $p_\mathrm{T}$, $v_n(p_\mathrm{T})$, $v_{0}(p_\mathrm{T})$, presented here, serves as a new probe that provides complementary insights into the system's properties. A Bayesian global analysis~\cite{Novak:2013bqa, Bernhard:2019bmu, Nijs:2020ors, JETSCAPE:2020mzn, Parkkila:2021yha, Parkkila:2021tqq, Heffernan:2023utr, Virta:2024avu} combining $v_{0}(p_\mathrm{T})$ with other measurements is key to extracting transport properties and refining our understanding of the EOS of the QCD medium.

In summary, the first measurement of $v_{0}(p_\mathrm{T})$, a novel observable that quantifies event-by-event fluctuations of radial flow and thereby constrains the collective radial expansion in heavy-ion collisions, is reported. The measured $v_{0}(p_\mathrm{T})$ evaluated using a pseudorapidity-gap technique to suppress nonflow contributions, captures long-range $p_\mathrm{T}$ correlations, analogous to $v_{n}(p_\mathrm{T})$ for azimuthal correlations. The results reveal characteristic mass ordering at low $p_\mathrm{T}$, consistent with hydrodynamic collective flow. At higher $p_\mathrm{T}$, protons exhibit larger values than pions and kaons, similar to the baryon-meson splitting in $v_2(p_\mathrm{T})$ and $v_3(p_\mathrm{T})$, suggesting quark recombination as the dominant particle-production mechanism. The hydrodynamic model of IP-Glasma+MUSIC+UrQMD describes the data well for $p_\mathrm{T} \lesssim$~2 GeV/$c$ across all centralities, with deviations observed for larger values of $p_\mathrm{T}$. In peripheral collisions, the consistency with HIJING suggests that hard scatterings and jet production play a more dominant role. Further comparison reveals that $v_{0}(p_\mathrm{T})$ is sensitive to the $\zeta/s$ and the EOS, while being relatively insensitive to the $\eta/s$, of the QCD medium formed in heavy-ion collisions. This sensitivity arises because both $\zeta/s$ and $c_s^{2}$ modify the system's isotropic expansion rate, affecting the underlying momentum-space correlations. In addition, $v_{0}(p_\mathrm{T})$ is sensitive to initial-state geometry and its fluctuations, which influence the build-up of pressure driving the collective radial expansion.


\newenvironment{acknowledgement}{\relax}{\relax}
\begin{acknowledgement}
\section*{Acknowledgements}

The ALICE Collaboration would like to thank all its engineers and technicians for their invaluable contributions to the construction of the experiment and the CERN accelerator teams for the outstanding performance of the LHC complex.
The ALICE Collaboration gratefully acknowledges the resources and support provided by all Grid centres and the Worldwide LHC Computing Grid (WLCG) collaboration.
The ALICE Collaboration acknowledges the following funding agencies for their support in building and running the ALICE detector:
A. I. Alikhanyan National Science Laboratory (Yerevan Physics Institute) Foundation (ANSL), State Committee of Science and World Federation of Scientists (WFS), Armenia;
Austrian Academy of Sciences, Austrian Science Fund (FWF): [M 2467-N36] and Nationalstiftung f\"{u}r Forschung, Technologie und Entwicklung, Austria;
Ministry of Communications and High Technologies, National Nuclear Research Center, Azerbaijan;
Conselho Nacional de Desenvolvimento Cient\'{\i}fico e Tecnol\'{o}gico (CNPq), Financiadora de Estudos e Projetos (Finep), Funda\c{c}\~{a}o de Amparo \`{a} Pesquisa do Estado de S\~{a}o Paulo (FAPESP) and Universidade Federal do Rio Grande do Sul (UFRGS), Brazil;
Bulgarian Ministry of Education and Science, within the National Roadmap for Research Infrastructures 2020-2027 (object CERN), Bulgaria;
Ministry of Education of China (MOEC) , Ministry of Science \& Technology of China (MSTC) and National Natural Science Foundation of China (NSFC), China;
Ministry of Science and Education and Croatian Science Foundation, Croatia;
Centro de Aplicaciones Tecnol\'{o}gicas y Desarrollo Nuclear (CEADEN), Cubaenerg\'{\i}a, Cuba;
Ministry of Education, Youth and Sports of the Czech Republic, Czech Republic;
The Danish Council for Independent Research | Natural Sciences, the VILLUM FONDEN and Danish National Research Foundation (DNRF), Denmark;
Helsinki Institute of Physics (HIP), Finland;
Commissariat \`{a} l'Energie Atomique (CEA) and Institut National de Physique Nucl\'{e}aire et de Physique des Particules (IN2P3) and Centre National de la Recherche Scientifique (CNRS), France;
Bundesministerium f\"{u}r Bildung und Forschung (BMBF) and GSI Helmholtzzentrum f\"{u}r Schwerionenforschung GmbH, Germany;
General Secretariat for Research and Technology, Ministry of Education, Research and Religions, Greece;
National Research, Development and Innovation Office, Hungary;
Department of Atomic Energy Government of India (DAE), Department of Science and Technology, Government of India (DST), University Grants Commission, Government of India (UGC) and Council of Scientific and Industrial Research (CSIR), India;
National Research and Innovation Agency - BRIN, Indonesia;
Istituto Nazionale di Fisica Nucleare (INFN), Italy;
Japanese Ministry of Education, Culture, Sports, Science and Technology (MEXT) and Japan Society for the Promotion of Science (JSPS) KAKENHI, Japan;
Consejo Nacional de Ciencia (CONACYT) y Tecnolog\'{i}a, through Fondo de Cooperaci\'{o}n Internacional en Ciencia y Tecnolog\'{i}a (FONCICYT) and Direcci\'{o}n General de Asuntos del Personal Academico (DGAPA), Mexico;
Nederlandse Organisatie voor Wetenschappelijk Onderzoek (NWO), Netherlands;
The Research Council of Norway, Norway;
Pontificia Universidad Cat\'{o}lica del Per\'{u}, Peru;
Ministry of Science and Higher Education, National Science Centre and WUT ID-UB, Poland;
Korea Institute of Science and Technology Information and National Research Foundation of Korea (NRF), Republic of Korea;
Ministry of Education and Scientific Research, Institute of Atomic Physics, Ministry of Research and Innovation and Institute of Atomic Physics and Universitatea Nationala de Stiinta si Tehnologie Politehnica Bucuresti, Romania;
Ministerstvo skolstva, vyskumu, vyvoja a mladeze SR, Slovakia;
National Research Foundation of South Africa, South Africa;
Swedish Research Council (VR) and Knut \& Alice Wallenberg Foundation (KAW), Sweden;
European Organization for Nuclear Research, Switzerland;
Suranaree University of Technology (SUT), National Science and Technology Development Agency (NSTDA) and National Science, Research and Innovation Fund (NSRF via PMU-B B05F650021), Thailand;
Turkish Energy, Nuclear and Mineral Research Agency (TENMAK), Turkey;
National Academy of  Sciences of Ukraine, Ukraine;
Science and Technology Facilities Council (STFC), United Kingdom;
National Science Foundation of the United States of America (NSF) and United States Department of Energy, Office of Nuclear Physics (DOE NP), United States of America.
In addition, individual groups or members have received support from:
Czech Science Foundation (grant no. 23-07499S), Czech Republic;
FORTE project, reg.\ no.\ CZ.02.01.01/00/22\_008/0004632, Czech Republic, co-funded by the European Union, Czech Republic;
European Research Council (grant no. 950692), European Union;
Deutsche Forschungs Gemeinschaft (DFG, German Research Foundation) ``Neutrinos and Dark Matter in Astro- and Particle Physics'' (grant no. SFB 1258), Germany;
ICSC - National Research Center for High Performance Computing, Big Data and Quantum Computing and FAIR - Future Artificial Intelligence Research, funded by the NextGenerationEU program (Italy).

\end{acknowledgement}

\bibliographystyle{utphys}   
\bibliography{bibliography2}

\newpage
\appendix

\section{Supplementary material}
\label{sec:Supp}

\subsection{\texorpdfstring{$v_{0}(p_\mathrm{T})$}{v0(p)T} and the radial-flow parameter \texorpdfstring{$\langle\beta_\mathrm{T}\rangle$}{beta(T)} }
\label{sec:blast}

\begin{figure}[ht]
  \begin{center}
  \includegraphics[width=\linewidth]{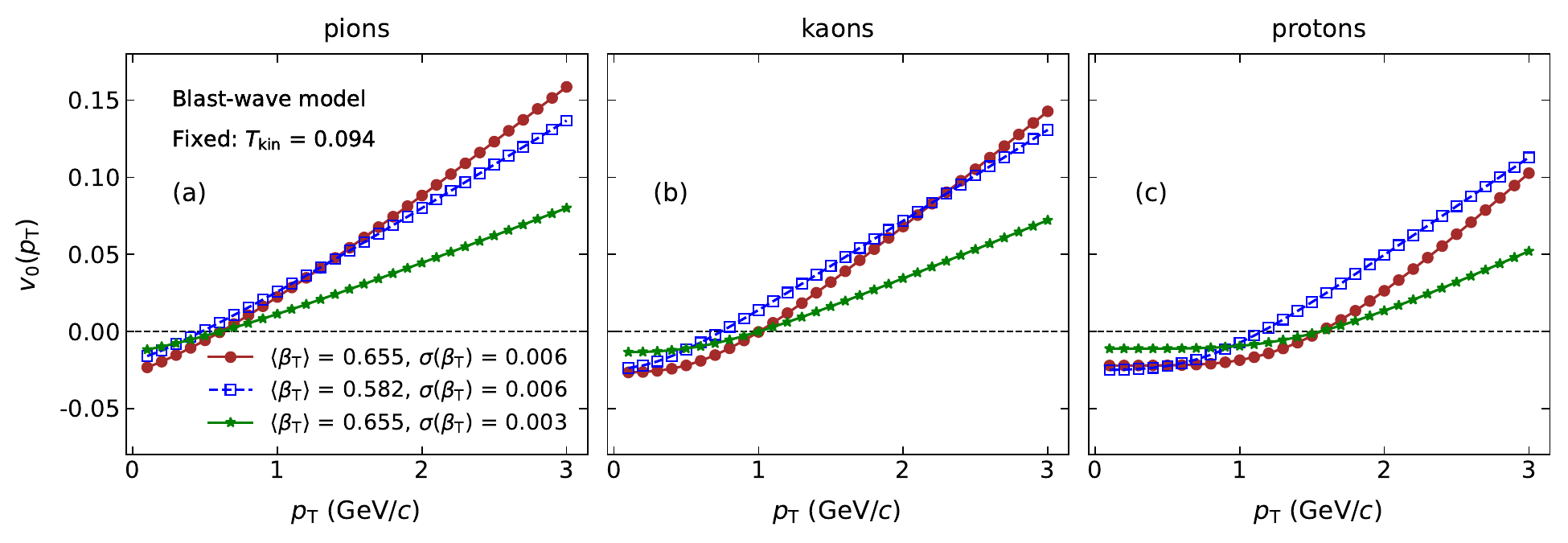}
  \end{center}
  \caption{$v_{0}(p_\mathrm{T})$ of pions (a), kaons (b), and protons (c) shown as a function of $p_\mathrm{T}$ using blast-wave model parameters from Ref.~\cite{ALICE:2019hno}. The open marker represents results for a slightly smaller value of $\langle\beta_\mathrm{T}\rangle$.  }
  \label{fig:v0pT_blast}
\end{figure}

This section demonstrates the connection between $v_{0}(p_\mathrm{T})$ and the $p_\mathrm{T}$-integrated radial-flow parameter $\langle\beta_\mathrm{T}\rangle$ using the Boltzmann-Gibbs blast-wave model~\cite{Schnedermann:1993ws}. While $v_{0}(p_\mathrm{T})$ is an event-by-event observable, experimental $p_\mathrm{T}$ spectra are typically measured over many events. However, fluctuations in $\langle\beta_\mathrm{T}\rangle$ can induce event-by-event variations in $p_\mathrm{T}$ spectra, influencing $v_{0}(p_\mathrm{T})$. To investigate this, $v_{0}(p_\mathrm{T})$ is computed by fixing $\langle\beta_\mathrm{T}\rangle$ and kinetic freeze-out temperature $T_\mathrm{kin}$, using parameters from Ref.~\cite{ALICE:2019hno}. The fluctuations in $\beta_\mathrm{T}$ are assumed to be Gaussian with a width of $\sigma(\beta_\mathrm{T})=0.006$. Figure~\ref{fig:v0pT_blast} presents the resulting $v_{0}(p_\mathrm{T})$ for pions, kaons, and protons at two different values of $\langle\beta_\mathrm{T}\rangle$. The blast-wave model calculations qualitatively capture the $p_\mathrm{T}$ dependence observed in experimental data, and the sign change arises from the correlation between fluctuations in radial flow and $p_\mathrm{T}$ spectra. It may be noted that in the absence of radial-flow fluctuations, $v_{0}(p_\mathrm{T})$ would be zero. Reducing $\sigma(\beta_\mathrm{T})$ to 0.003 significantly modifies $v_{0}(p_\mathrm{T})$, highlighting its sensitivity to the strength of radial flow fluctuations. The observed nonzero values in experimental data suggest the presence of such fluctuations, with $\langle\beta_\mathrm{T}\rangle$ influencing the $p_\mathrm{T}$ dependence of $v_{0}(p_\mathrm{T})$. A detailed study reported in Ref.~\cite{Saha:2025nyu} further extends this analysis by demonstrating that, although $v_{0}(p_\mathrm{T})$ is primarily driven by radial flow, the observable also exhibits sensitivity to $T_\mathrm{kin}$ and to the corresponding width of event-by-event fluctuation, $\sigma(T_\mathrm{kin})$.

\subsection{Effect of \texorpdfstring{$\Delta\eta$}{eta}-gap variation}
\label{sec:etadep}

\begin{figure}[ht]
  \begin{center}
    \includegraphics[width=0.32\linewidth]{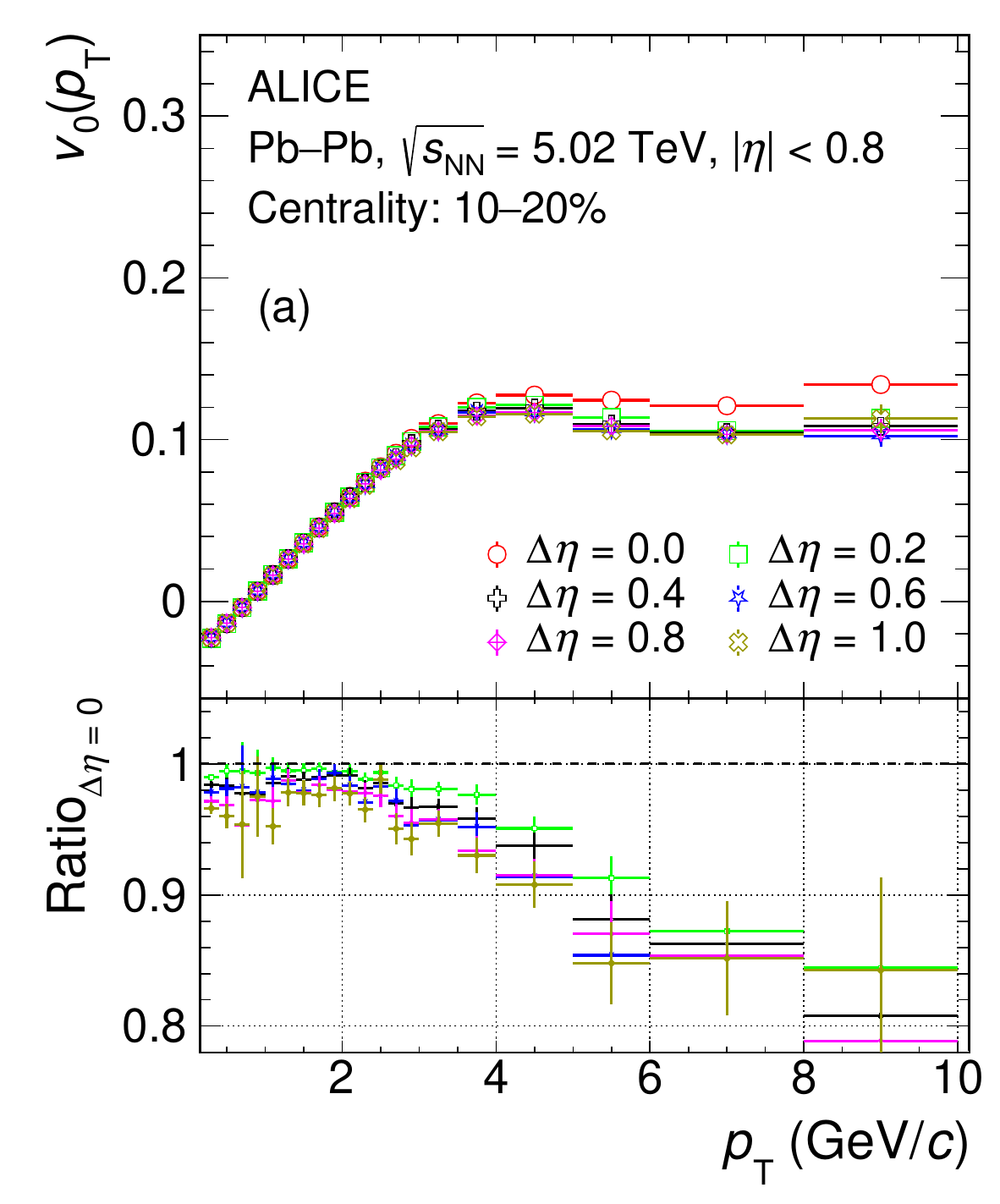}
    \includegraphics[width=0.32\linewidth]{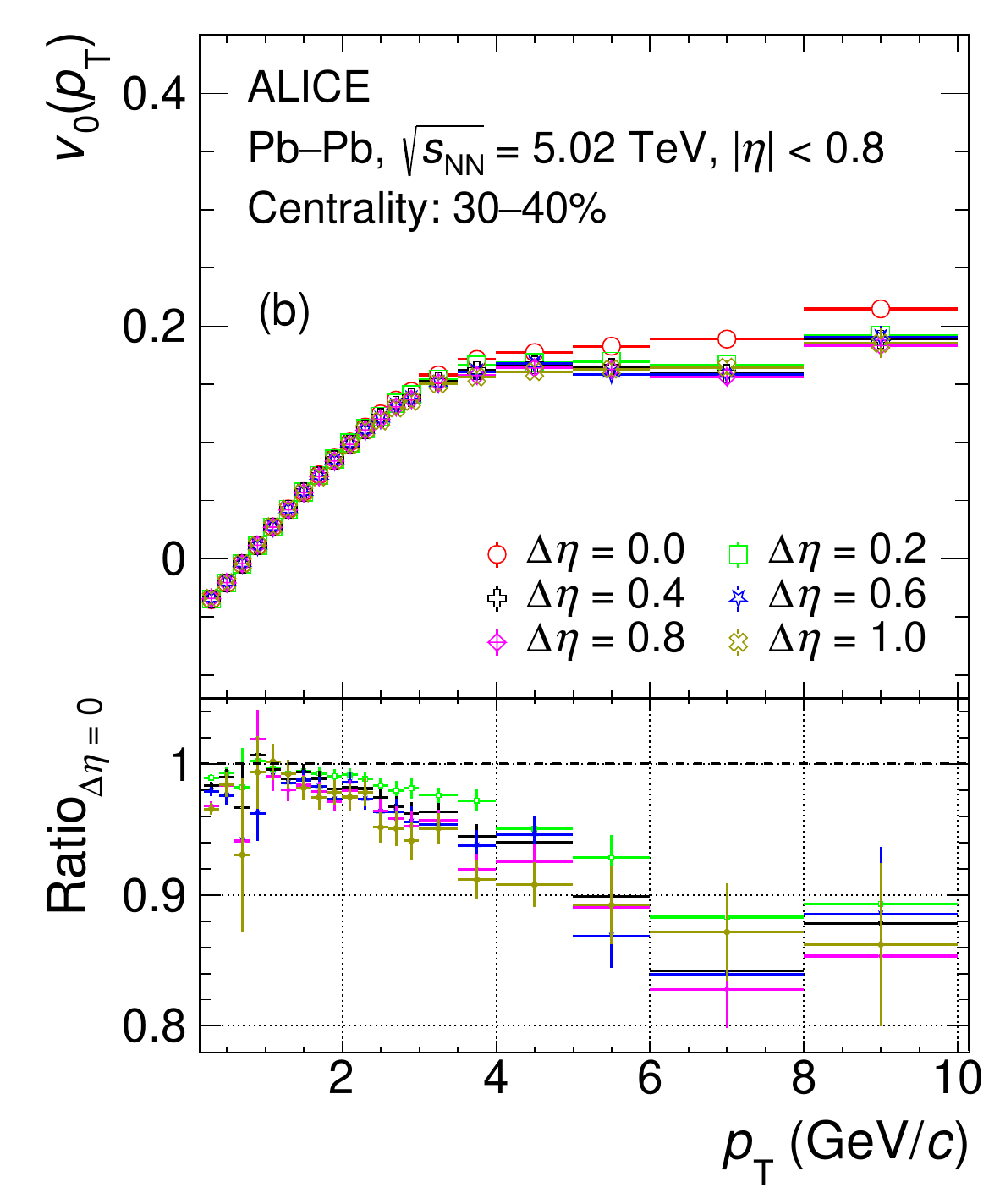}
    \includegraphics[width=0.32\linewidth]{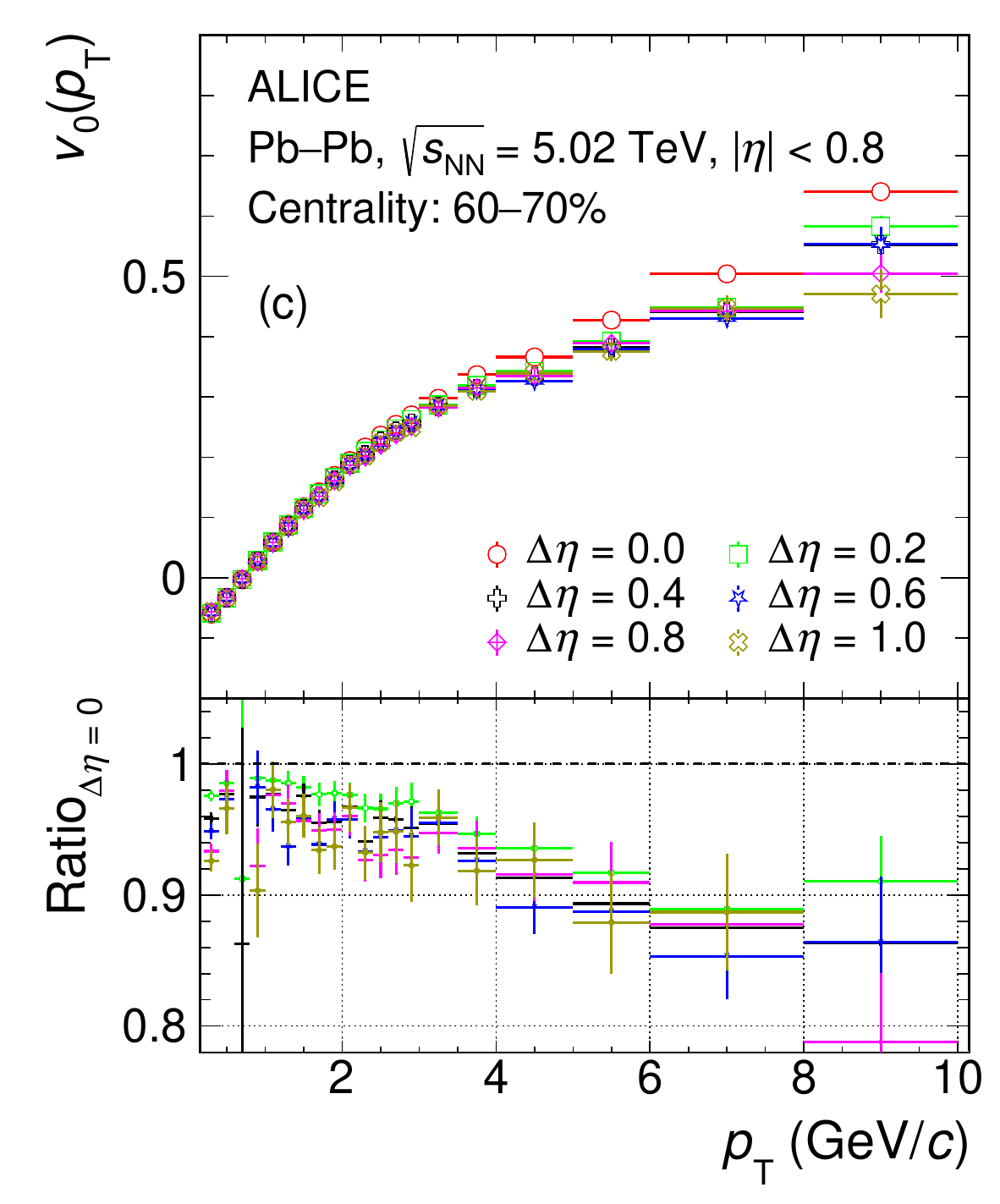}
  \end{center}
  \caption{$v_{0}(p_\mathrm{T})$ of inclusive charged particles shown as a function of $p_\mathrm{T}$ for centrality intervals 10$-$20\% (a), 30$-$40\% (b), and 60$-$70\% (c) for varying pseudorapidity gap ($\Delta\eta$) in Pb$-$Pb collisions at $\sqrt{s_\mathrm{NN}} = 5.02$ TeV. The error bars represent statistical uncertainties. The bottom panel presents the ratio relative to the results for $\Delta\eta = 0$.}
  \label{fig:v0pT_etagap}
\end{figure}

$v_{0}(p_\mathrm{T})$ of inclusive charged particles is studied as a function of $p_\mathrm{T}$ for varying $\Delta\eta$ gaps (0--1, in steps of 0.2) across different centralities, as shown in Fig.~\ref{fig:v0pT_etagap}. The goal is to investigate the influence of short-range correlations, or nonflow effects, which may arise from resonance decays, near-side jets, and other few-particle correlations. For $p_\mathrm{T} < 3$~GeV/$c$, variations in $v_{0}(p_\mathrm{T})$ with $\Delta\eta$ relative to $\Delta\eta=0$, remain below 2\% in central and semicentral collisions and 8\% in peripheral collisions. At higher $p_\mathrm{T}$, a difference of up to $\sim$15\% between results with and without a $\Delta\eta$ gap suggests a modest influence from short-range correlations. The results remain stable for $\Delta\eta > 0.2$, leading to the choice of $\Delta\eta=0.4$ as the optimal gap, with variations ($\Delta\eta=0.5$ and 0.6) included in the systematic uncertainty estimates.

\subsection{MC closure test}
\label{sec:mc_closure}
\begin{figure}[ht]
  \begin{center}
    \includegraphics[width=0.32\linewidth]{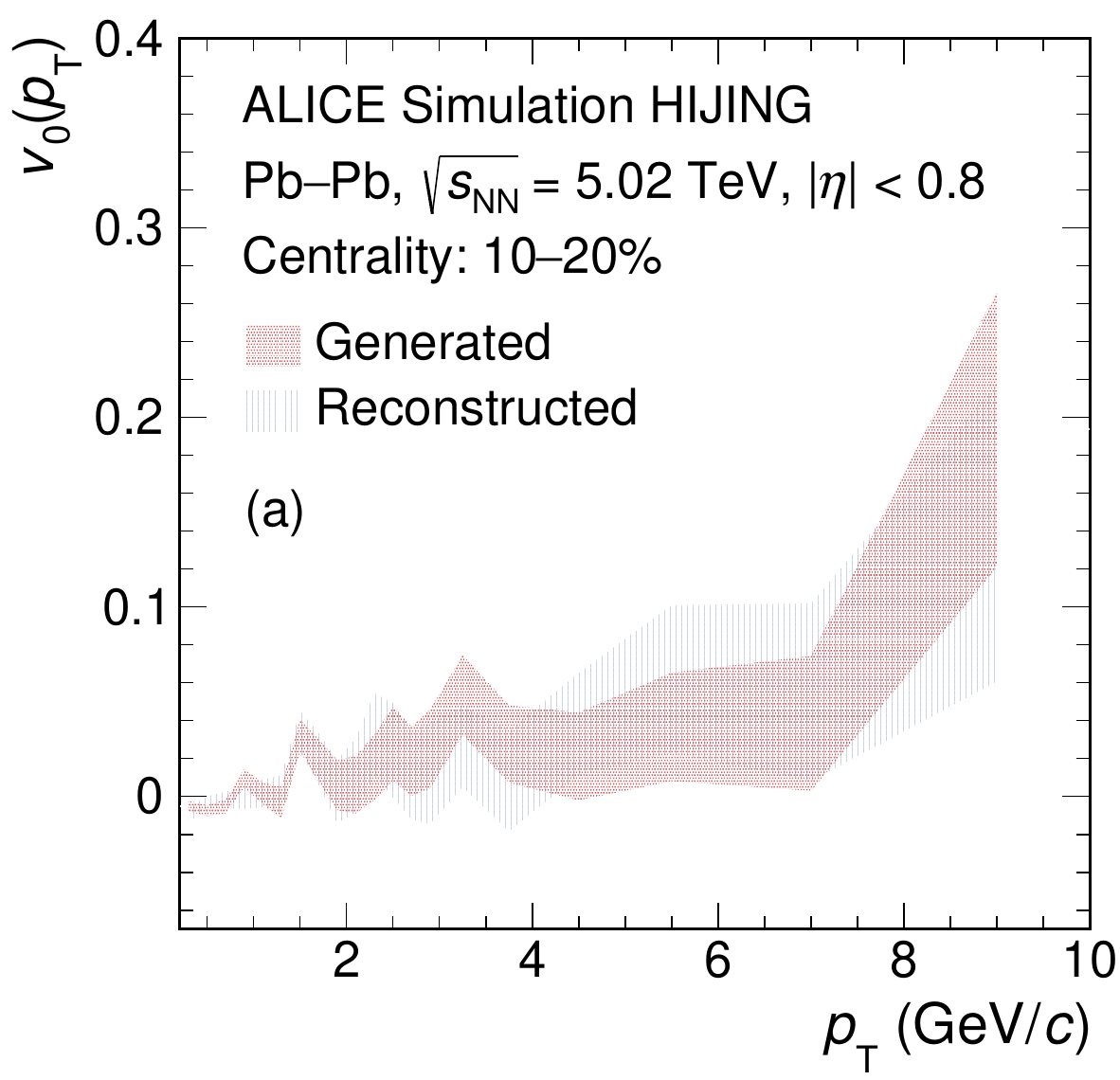}
    \includegraphics[width=0.32\linewidth]{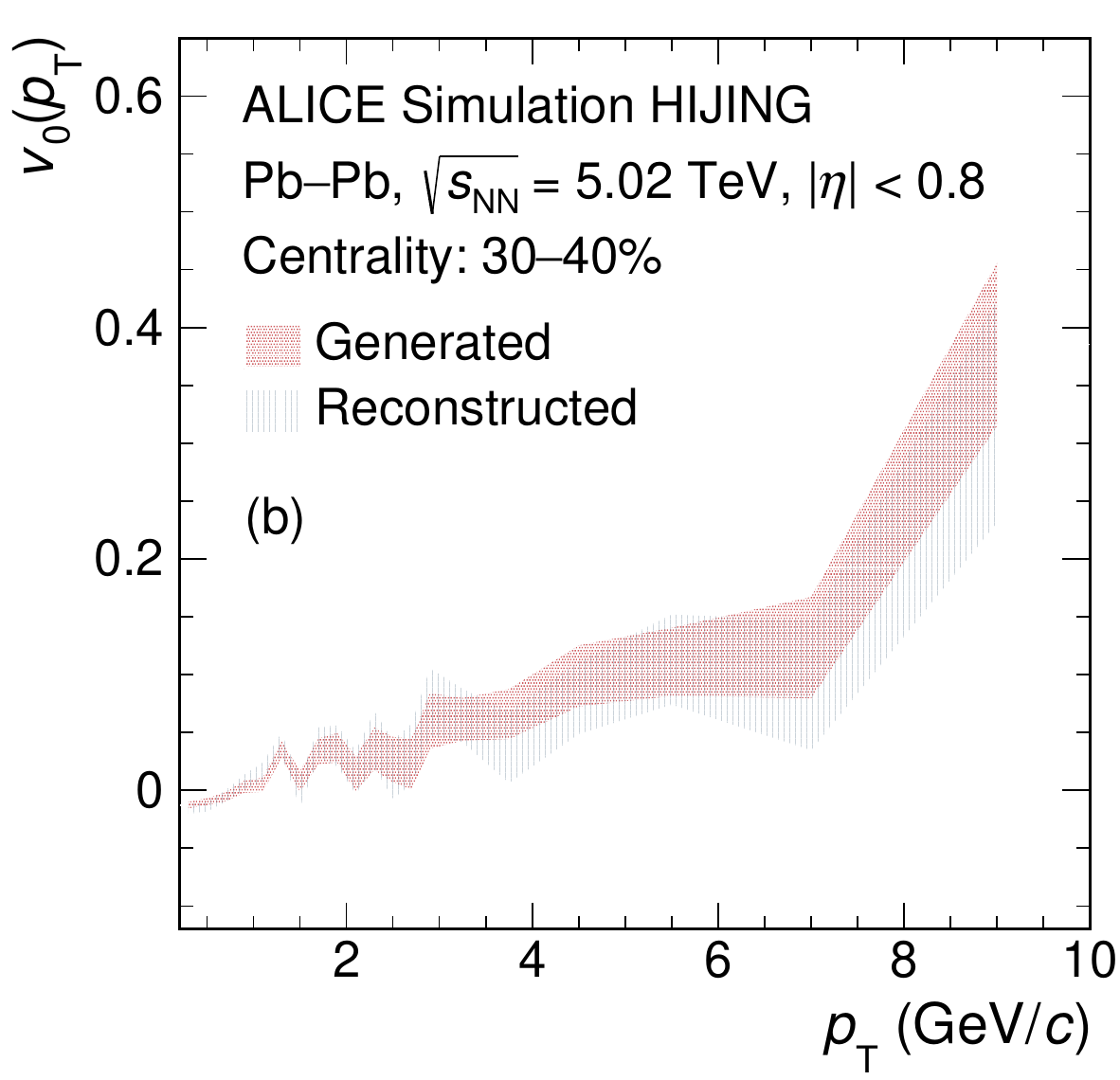}
    \includegraphics[width=0.32\linewidth]{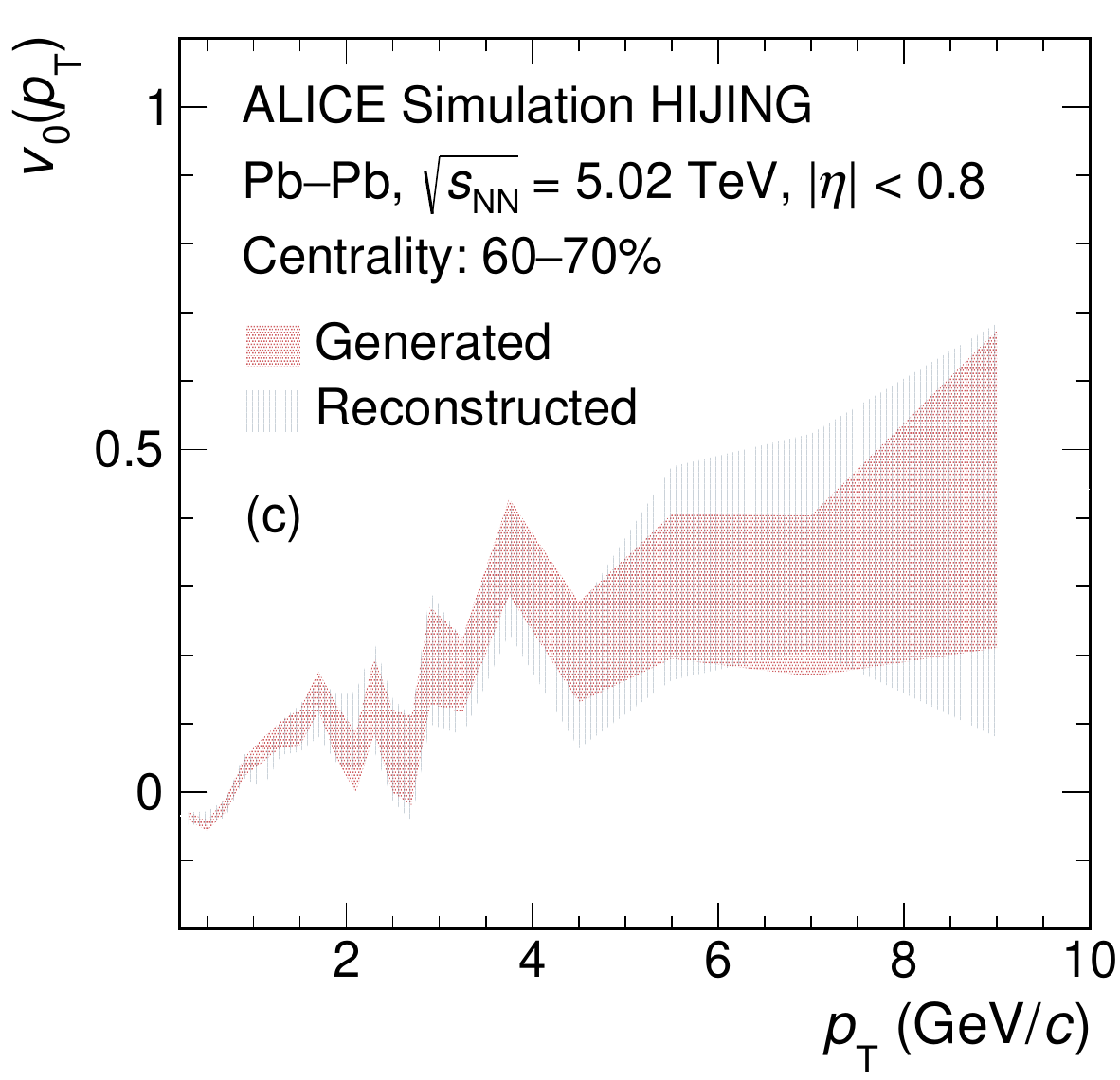}
  \end{center}
  \caption{HIJING model based calculations of $v_{0}(p_\mathrm{T})$ of inclusive charged particles as a function of $p_\mathrm{T}$ for centrality intervals 10$-$20\% (a), 30$-$40\% (b), and 60$-$70\% (c) in Pb$-$Pb collisions at $\sqrt{s_\mathrm{NN}} = 5.02$ TeV. The results at the generated and reconstructed level are shown, with lines connecting the central values and bands representing the statistical uncertainties.}
  \label{fig:v0pT_MCclosure}
\end{figure}

The MC closure test uses HIJING model~\cite{Wang:1991hta, ToporPop:2004lve} to generate events and GEANT3~\cite{Brun:1994aa} simulations for particle transport through the ALICE detector geometry. The transported particles are reconstructed using the same procedure as experimental data. Figure~\ref{fig:v0pT_MCclosure} shows the comparison of $v_{0}(p_\mathrm{T})$ obtained from the generated events with those obtained from the corresponding reconstructed events (without applying efficiency corrections) for inclusive charged particles across different centrality intervals. The generated results are in agreement with the reconstructed results within uncertainties, confirming the absence of significant detector efficiency effects on the observable. Although statistical uncertainties remain sizable, the observed correlation between fluctuations in generated and reconstructed $v_{0}(p_\mathrm{T})$ provides additional confidence in the closure test.

\subsection{Systematic uncertainty}
\label{sec:app_sys}

The systematic uncertainties from various sources listed in Table~\ref{tab:sys} are expressed as relative uncertainties with respect to the central values of the data points, and averaged over all $p_\mathrm{T}$ bins. These contributions vary by particle species and centrality intervals, and exhibit a general increase with $p_\mathrm{T}$. The detailed breakdown of the systematic uncertainties for 30$-$40\% centrality class is presented as an example in Table~\ref{tab:sys}. In this centrality interval, the total uncertainties (obtained by adding individual sources in quadrature) amount to 3.8\%, 18.9\%, 11.1\%, and 15.1\% for $\mathrm{h}^{\pm}$, $\pi^{\pm}$, $\mathrm{K}^{\pm}$, and $\mathrm{p}(\bar{\mathrm{p}})$, respectively.

\begin{table}[ht]
  \begin{center}
      \caption{ \label{tab:sys} Contributions to the total systematic uncertainty of $v_{0}(p_\mathrm{T})$ are shown for inclusive charged particles ($\mathrm{h}^{\pm}$), pions ($\pi^{\pm}$), kaons ($\mathrm{K}^{\pm}$), and protons ($\mathrm{p}(\bar{\mathrm{p}})$) in Pb$-$Pb collisions at $\sqrt{s_\mathrm{NN}} = 5.02$ TeV for the 30$-$40\% centrality interval. The values represent the average uncertainty over all $p_\mathrm{T}$ bins.}

      \begin{tabular}{lcccc}\\ \hline
      Sources of systematic uncertainty & $\mathrm{h}^{\pm}$ & $\pi^{\pm}$ & $\mathrm{K}^{\pm}$ & $\mathrm{p}(\bar{\mathrm{p}}$)\\ \hline
      Primary vertex        & 0.5\% & 0.5\%  & 0.6\%  & 1.1\%  \\
      Pileup rejection       & 0.3\%  & 0.5\%  & 0.9\%  & 1.1\%  \\
      Centrality estimation & 1.3\% & 2.1\% & 2.2\% & 2.7\% \\
      DCA                   & 2.8\% & 1.5\% & 5.2\% & 12.8\% \\
      TPC crossed rows      & 1.2\% & 0.6\% & 1.5\% & 2.3\% \\
      TPC $\chi^{2}$ fit     & 0.5\% & 0.4\% & 0.6\% & 1.2\% \\
      ITS $\chi^{2}$ fit   &  0.3\% & 0.3\% & 0.7\% & 0.9\% \\
      $\Delta\eta$ gap            & 1.1\% & 1.4\% & 3.5\% & 3.4\% \\
      Particle identification     & --     & 18.2\% & 7.9\% & 2.8\% \\\hline
      Total & 3.8\% & 18.9\% & 11.1\% & 15.1\% \\
      \hline
      \end{tabular}
  \end{center}
\end{table}

\subsection{Effect of \texorpdfstring{$\varphi$}{phi} acceptance}
\label{sec:phidep}
\begin{figure}[hb]
  \begin{center}
    \includegraphics[width=0.32\linewidth]{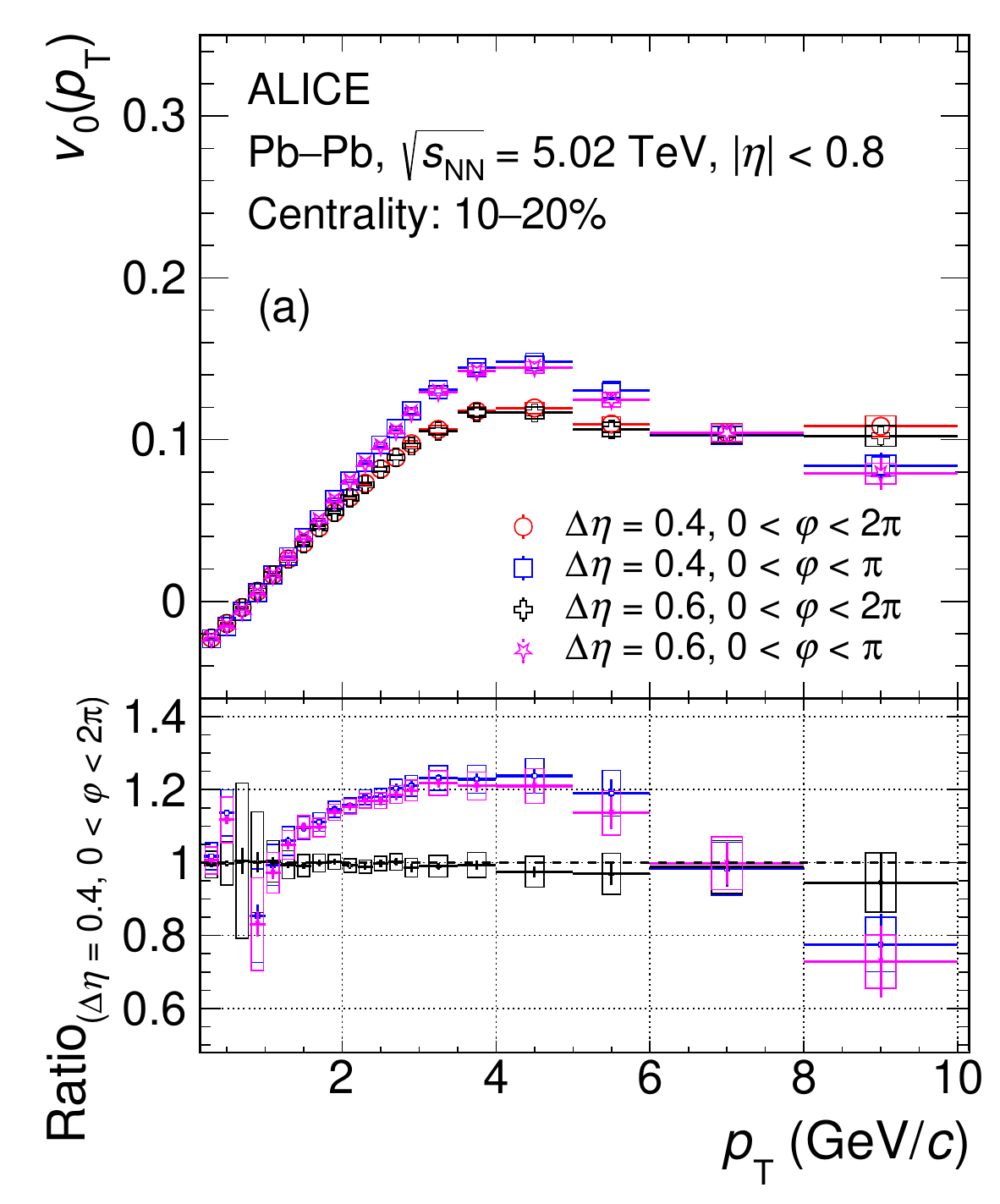}
    \includegraphics[width=0.32\linewidth]{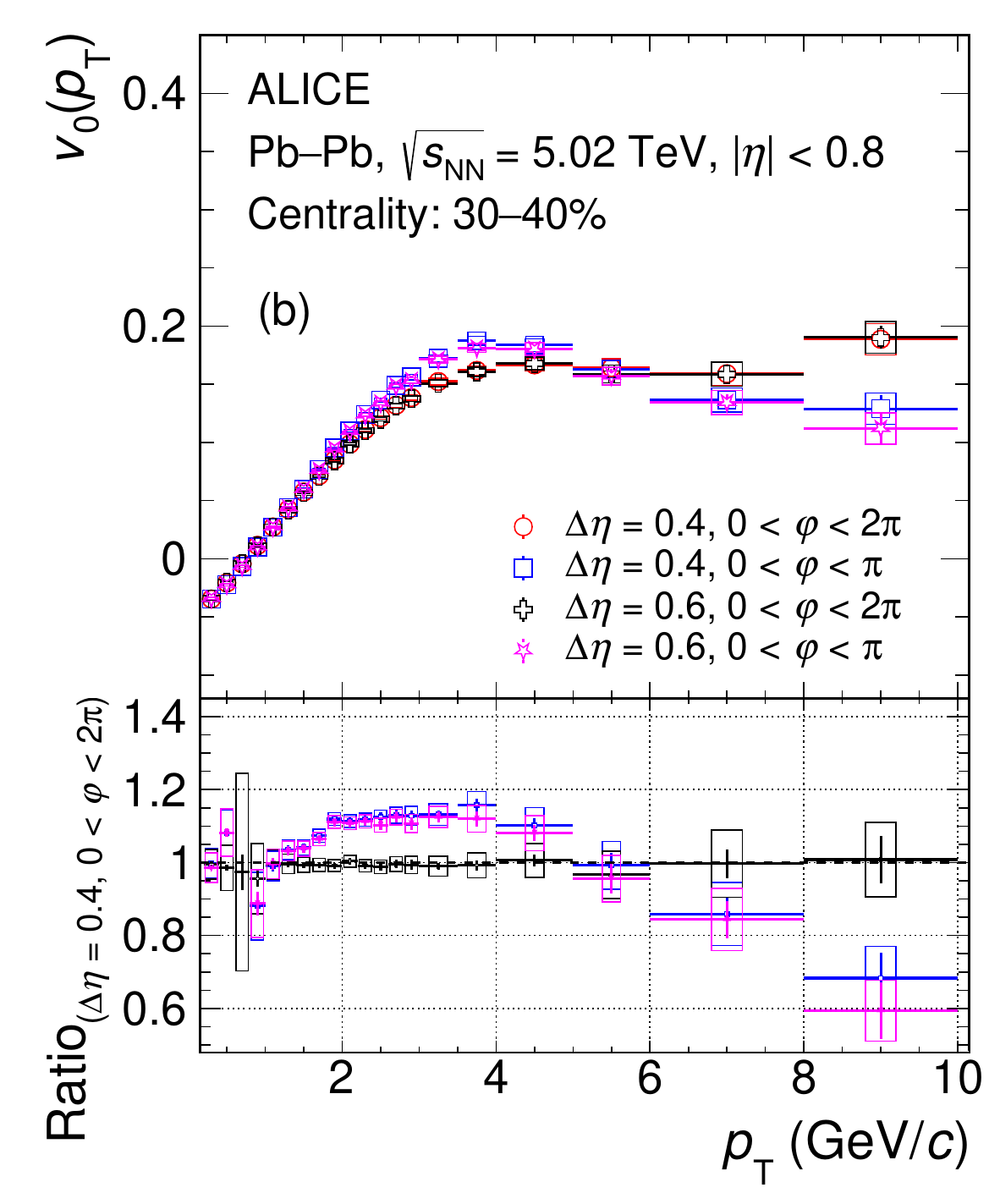}
    \includegraphics[width=0.32\linewidth]{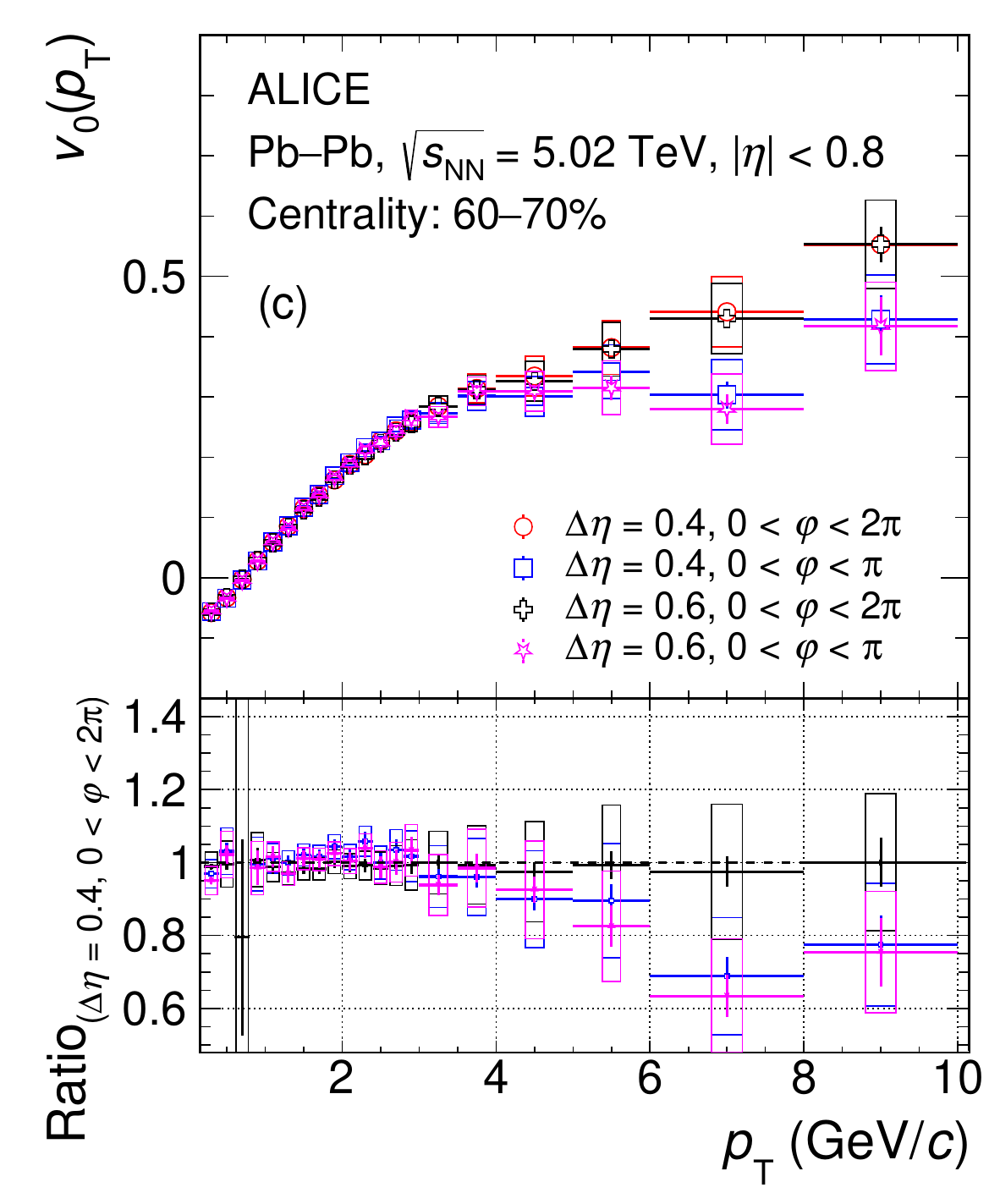}
  \end{center}
  \caption{$v_{0}(p_\mathrm{T})$ of inclusive charged particles shown as a function of $p_\mathrm{T}$ for centrality intervals 10$-$20\% (a), 30$-$40\% (b), and 60$-$70\% (c) for varying pseudorapidity gap ($\Delta\eta$) and azimuthal acceptance ($\varphi$) in Pb$-$Pb collisions at $\sqrt{s_\mathrm{NN}} = 5.02$ TeV. The statistical (systematic) uncertainties are represented by vertical bars (boxes). The bottom panel presents the ratio relative to the results for $\Delta\eta = 0.4, 0 < \varphi < 2\pi$.}
  \label{fig:v0pT_etagap_phi}
\end{figure}

Nonflow contributions from back-to-back dijets, which could extend over long ranges in $\eta$, may not be suppressed by applying a $\Delta\eta$ gap. To assess their impact, the $\varphi$ acceptance is restricted in 0 to $\pi$ and compared with the full acceptance (0 to 2$\pi$). Figure~\ref{fig:v0pT_etagap_phi} presents $v_{0}(p_\mathrm{T})$ of inclusive charged particles for different $\Delta\eta$ and $\varphi$-acceptances across centrality intervals. At low $p_\mathrm{T}$ ($< 1.5$~GeV/$c$), all configurations yield similar results, suggesting minimal effect of the nonflow contributions in this region. For $p_\mathrm{T}> 1.5$~GeV/$c$, a maximum variation of $\sim$20\% is observed between full-$\varphi$ and half-$\varphi$ acceptance across the centrality intervals, indicating possible influence of dijet-like correlations. The observed deviations also reveal a $p_\mathrm{T}$ dependence, that varies from central to peripheral collisions.

\subsection{NCQ scaling of \texorpdfstring{$v_{0}(p_\mathrm{T})$}{v0(pt)}}
\label{sec:NCQ}
\begin{figure}[h]
  \begin{center}
    \includegraphics[width=\linewidth]{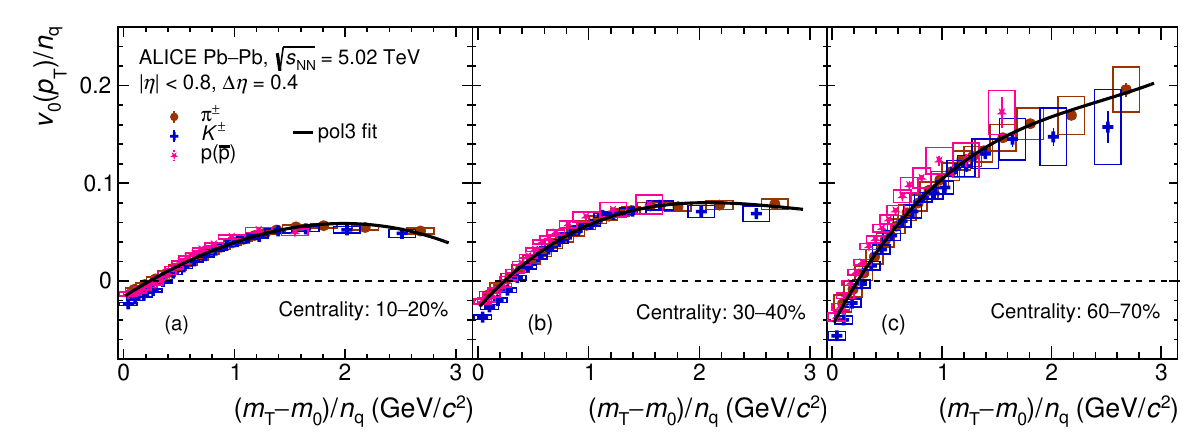}
  \end{center}
  \caption{$v_{0}(p_\mathrm{T})$ of pions ($\pi^{\pm}$), kaons ($\mathrm{K}^{\pm}$), and protons ($\mathrm{p}(\bar{\mathrm{p}})$) scaled by number of constituent quarks ($n_q$) shown as a function of transverse kinetic energy per constituent quark, $(m_\mathrm{T}-m_{0})/n_{q}$ in Pb$-$Pb collisions at $\sqrt{s_\mathrm{NN}} = 5.02$ TeV for centrality intervals 10$-$20\% (a), 30$-$40\% (b), and 60$-$70\% (c). The line represents a third-order polynomial function fitted to the pion data.}
  \label{fig:v0pT_piKprotNCQ}
\end{figure}

The behavior of $v_{0}(p_\mathrm{T})$ for identified hadrons is further explored in Fig.~\ref{fig:v0pT_piKprotNCQ}. The observable, normalized by the number of constituent quarks ($n_q$), is plotted against transverse kinetic energy per quark, $(m_\mathrm{T}-m_{0})/n_{q}$, where $m_\mathrm{T} = \sqrt{p_\mathrm{T}^2+m_{0}^2}$, and $m_{0}$ denotes the particle rest mass. After scaling, the results for pions, kaons, and protons display a clear tendency to follow a common trend across different centralities, indicating an approximate NCQ scaling. Such scaling behavior has traditionally been regarded as evidence for collective motion emerging at the partonic stage. While the scaling is not exact over the full measured range and small deviations exist at lower values of $(m_\mathrm{T}-m_{0})/n_{q}$, the overall consistency aligns well with expectations from quark recombination models. This qualitative agreement supports the interpretation that quark‑level collectivity contributes to the observed dynamics, although hadronic effects may still influence the degree of scaling.

\newpage
\section{The ALICE Collaboration}
\label{app:collab}
\begin{flushleft} 
\small

S.~Acharya\,\orcidlink{0000-0002-9213-5329}\,$^{\rm 50}$, 
G.~Aglieri Rinella\,\orcidlink{0000-0002-9611-3696}\,$^{\rm 32}$, 
L.~Aglietta\,\orcidlink{0009-0003-0763-6802}\,$^{\rm 24}$, 
M.~Agnello\,\orcidlink{0000-0002-0760-5075}\,$^{\rm 29}$, 
N.~Agrawal\,\orcidlink{0000-0003-0348-9836}\,$^{\rm 25}$, 
Z.~Ahammed\,\orcidlink{0000-0001-5241-7412}\,$^{\rm 133}$, 
S.~Ahmad\,\orcidlink{0000-0003-0497-5705}\,$^{\rm 15}$, 
S.U.~Ahn\,\orcidlink{0000-0001-8847-489X}\,$^{\rm 71}$, 
I.~Ahuja\,\orcidlink{0000-0002-4417-1392}\,$^{\rm 36}$, 
Z.~Akbar$^{\rm 81}$
% ...

\section*{Affiliation Notes}

$^{\rm I}$ Deceased\\
$^{\rm II}$ Also at: Max-Planck-Institut fur Physik, Munich, Germany\\
$^{\rm III}$ Also at: Italian National Agency for New Technologies, Energy and Sustainable Economic Development (ENEA), Bologna, Italy\\
$^{\rm IV}$ Also at: Instituto de Fisica da Universidade de Sao Paulo\\
$^{\rm V}$ Also at: Dipartimento DET del Politecnico di Torino, Turin, Italy\\
$^{\rm VI}$ Also at: Department of Applied Physics, Aligarh Muslim University, Aligarh, India\\
$^{\rm VII}$ Also at: Institute of Theoretical Physics, University of Wroclaw, Poland\\
$^{\rm VIII}$ Also at: Facultad de Ciencias, Universidad Nacional Aut\'{o}noma de M\'{e}xico, Mexico City, Mexico\\

\section*{Collaboration Institutes}

$^{1}$ A.I. Alikhanyan National Science Laboratory (Yerevan Physics Institute) Foundation, Yerevan, Armenia\\
$^{2}$ AGH University of Krakow, Cracow, Poland\\
$^{3}$ Bogolyubov Institute for Theoretical Physics, National Academy of Sciences of Ukraine, Kiev, Ukraine\\
$^{4}$ Bose Institute, Department of Physics  and Centre for Astroparticle Physics and Space Science (CAPSS), Kolkata, India\\
$^{5}$ California Polytechnic State University, San Luis Obispo, California, United States\\
$^{6}$ Central China Normal University, Wuhan, China\\
$^{7}$ Centro de Aplicaciones Tecnol\'{o}gicas y Desarrollo Nuclear (CEADEN), Havana, Cuba\\
$^{8}$ Centro de Investigaci\'{o}n y de Estudios Avanzados (CINVESTAV), Mexico City and M\'{e}rida, Mexico\\
$^{9}$ Chicago State University, Chicago, Illinois, United States\\
$^{10}$ China Nuclear Data Center, China Institute of Atomic Energy, Beijing, China\\
$^{11}$ China University of Geosciences, Wuhan, China\\
$^{12}$ Chungbuk National University, Cheongju, Republic of Korea\\
$^{13}$ Comenius University Bratislava, Faculty of Mathematics, Physics and Informatics, Bratislava, Slovak Republic\\
$^{14}$ Creighton University, Omaha, Nebraska, United States\\
$^{15}$ Department of Physics, Aligarh Muslim University, Aligarh, India\\
$^{16}$ Department of Physics, Pusan National University, Pusan, Republic of Korea\\
$^{17}$ Department of Physics, Sejong University, Seoul, Republic of Korea\\
$^{18}$ Department of Physics, University of California, Berkeley, California, United States\\
$^{19}$ Department of Physics, University of Oslo, Oslo, Norway\\
$^{20}$ Department of Physics and Technology, University of Bergen, Bergen, Norway\\
$^{21}$ Dipartimento di Fisica, Universit\`{a} di Pavia, Pavia, Italy\\
$^{22}$ Dipartimento di Fisica dell'Universit\`{a} and Sezione INFN, Cagliari, Italy\\
$^{23}$ Dipartimento di Fisica dell'Universit\`{a} and Sezione INFN, Trieste, Italy\\
$^{24}$ Dipartimento di Fisica dell'Universit\`{a} and Sezione INFN, Turin, Italy\\
$^{25}$ Dipartimento di Fisica e Astronomia dell'Universit\`{a} and Sezione INFN, Bologna, Italy\\
$^{26}$ Dipartimento di Fisica e Astronomia dell'Universit\`{a} and Sezione INFN, Catania, Italy\\
$^{27}$ Dipartimento di Fisica e Astronomia dell'Universit\`{a} and Sezione INFN, Padova, Italy\\
$^{28}$ Dipartimento di Fisica `E.R.~Caianiello' dell'Universit\`{a} and Gruppo Collegato INFN, Salerno, Italy\\
$^{29}$ Dipartimento DISAT del Politecnico and Sezione INFN, Turin, Italy\\
$^{30}$ Dipartimento di Scienze MIFT, Universit\`{a} di Messina, Messina, Italy\\
$^{31}$ Dipartimento Interateneo di Fisica `M.~Merlin' and Sezione INFN, Bari, Italy\\
$^{32}$ European Organization for Nuclear Research (CERN), Geneva, Switzerland\\
$^{33}$ Faculty of Electrical Engineering, Mechanical Engineering and Naval Architecture, University of Split, Split, Croatia\\
$^{34}$ Faculty of Nuclear Sciences and Physical Engineering, Czech Technical University in Prague, Prague, Czech Republic\\
$^{35}$ Faculty of Physics, Sofia University, Sofia, Bulgaria\\
$^{36}$ Faculty of Science, P.J.~\v{S}af\'{a}rik University, Ko\v{s}ice, Slovak Republic\\
$^{37}$ Faculty of Technology, Environmental and Social Sciences, Bergen, Norway\\
$^{38}$ Frankfurt Institute for Advanced Studies, Johann Wolfgang Goethe-Universit\"{a}t Frankfurt, Frankfurt, Germany\\
$^{39}$ Fudan University, Shanghai, China\\
$^{40}$ Gangneung-Wonju National University, Gangneung, Republic of Korea\\
$^{41}$ Gauhati University, Department of Physics, Guwahati, India\\
$^{42}$ Helmholtz-Institut f\"{u}r Strahlen- und Kernphysik, Rheinische Friedrich-Wilhelms-Universit\"{a}t Bonn, Bonn, Germany\\
$^{43}$ Helsinki Institute of Physics (HIP), Helsinki, Finland\\
$^{44}$ High Energy Physics Group,  Universidad Aut\'{o}noma de Puebla, Puebla, Mexico\\
$^{45}$ Horia Hulubei National Institute of Physics and Nuclear Engineering, Bucharest, Romania\\
$^{46}$ HUN-REN Wigner Research Centre for Physics, Budapest, Hungary\\
$^{47}$ Indian Institute of Technology Bombay (IIT), Mumbai, India\\
$^{48}$ Indian Institute of Technology Indore, Indore, India\\
$^{49}$ INFN, Laboratori Nazionali di Frascati, Frascati, Italy\\
$^{50}$ INFN, Sezione di Bari, Bari, Italy\\
$^{51}$ INFN, Sezione di Bologna, Bologna, Italy\\
$^{52}$ INFN, Sezione di Cagliari, Cagliari, Italy\\
$^{53}$ INFN, Sezione di Catania, Catania, Italy\\
$^{54}$ INFN, Sezione di Padova, Padova, Italy\\
$^{55}$ INFN, Sezione di Pavia, Pavia, Italy\\
$^{56}$ INFN, Sezione di Torino, Turin, Italy\\
$^{57}$ INFN, Sezione di Trieste, Trieste, Italy\\
$^{58}$ Inha University, Incheon, Republic of Korea\\
$^{59}$ Institute for Gravitational and Subatomic Physics (GRASP), Utrecht University/Nikhef, Utrecht, Netherlands\\
$^{60}$ Institute of Experimental Physics, Slovak Academy of Sciences, Ko\v{s}ice, Slovak Republic\\
$^{61}$ Institute of Physics, Homi Bhabha National Institute, Bhubaneswar, India\\
$^{62}$ Institute of Physics of the Czech Academy of Sciences, Prague, Czech Republic\\
$^{63}$ Institute of Space Science (ISS), Bucharest, Romania\\
$^{64}$ Institut f\"{u}r Kernphysik, Johann Wolfgang Goethe-Universit\"{a}t Frankfurt, Frankfurt, Germany\\
$^{65}$ Instituto de Ciencias Nucleares, Universidad Nacional Aut\'{o}noma de M\'{e}xico, Mexico City, Mexico\\
$^{66}$ Instituto de F\'{i}sica, Universidade Federal do Rio Grande do Sul (UFRGS), Porto Alegre, Brazil\\
$^{67}$ Instituto de F\'{\i}sica, Universidad Nacional Aut\'{o}noma de M\'{e}xico, Mexico City, Mexico\\
$^{68}$ iThemba LABS, National Research Foundation, Somerset West, South Africa\\
$^{69}$ Jeonbuk National University, Jeonju, Republic of Korea\\
$^{70}$ Johann-Wolfgang-Goethe Universit\"{a}t Frankfurt Institut f\"{u}r Informatik, Fachbereich Informatik und Mathematik, Frankfurt, Germany\\
$^{71}$ Korea Institute of Science and Technology Information, Daejeon, Republic of Korea\\
$^{72}$ Laboratoire de Physique Subatomique et de Cosmologie, Universit\'{e} Grenoble-Alpes, CNRS-IN2P3, Grenoble, France\\
$^{73}$ Lawrence Berkeley National Laboratory, Berkeley, California, United States\\
$^{74}$ Lund University Department of Physics, Division of Particle Physics, Lund, Sweden\\
$^{75}$ Nagasaki Institute of Applied Science, Nagasaki, Japan\\
$^{76}$ Nara Women{'}s University (NWU), Nara, Japan\\
$^{77}$ National and Kapodistrian University of Athens, School of Science, Department of Physics , Athens, Greece\\
$^{78}$ National Centre for Nuclear Research, Warsaw, Poland\\
$^{79}$ National Institute of Science Education and Research, Homi Bhabha National Institute, Jatni, India\\
$^{80}$ National Nuclear Research Center, Baku, Azerbaijan\\
$^{81}$ National Research and Innovation Agency - BRIN, Jakarta, Indonesia\\
$^{82}$ Niels Bohr Institute, University of Copenhagen, Copenhagen, Denmark\\
$^{83}$ Nikhef, National institute for subatomic physics, Amsterdam, Netherlands\\
$^{84}$ Nuclear Physics Group, STFC Daresbury Laboratory, Daresbury, United Kingdom\\
$^{85}$ Nuclear Physics Institute of the Czech Academy of Sciences, Husinec-\v{R}e\v{z}, Czech Republic\\
$^{86}$ Oak Ridge National Laboratory, Oak Ridge, Tennessee, United States\\
$^{87}$ Ohio State University, Columbus, Ohio, United States\\
$^{88}$ Physics department, Faculty of science, University of Zagreb, Zagreb, Croatia\\
$^{89}$ Physics Department, Panjab University, Chandigarh, India\\
$^{90}$ Physics Department, University of Jammu, Jammu, India\\
$^{91}$ Physics Program and International Institute for Sustainability with Knotted Chiral Meta Matter (WPI-SKCM$^{2}$), Hiroshima University, Hiroshima, Japan\\
$^{92}$ Physikalisches Institut, Eberhard-Karls-Universit\"{a}t T\"{u}bingen, T\"{u}bingen, Germany\\
$^{93}$ Physikalisches Institut, Ruprecht-Karls-Universit\"{a}t Heidelberg, Heidelberg, Germany\\
$^{94}$ Physik Department, Technische Universit\"{a}t M\"{u}nchen, Munich, Germany\\
$^{95}$ Politecnico di Bari and Sezione INFN, Bari, Italy\\
$^{96}$ Research Division and ExtreMe Matter Institute EMMI, GSI Helmholtzzentrum f\"ur Schwerionenforschung GmbH, Darmstadt, Germany\\
$^{97}$ Saga University, Saga, Japan\\
$^{98}$ Saha Institute of Nuclear Physics, Homi Bhabha National Institute, Kolkata, India\\
$^{99}$ School of Physics and Astronomy, University of Birmingham, Birmingham, United Kingdom\\
$^{100}$ Secci\'{o}n F\'{\i}sica, Departamento de Ciencias, Pontificia Universidad Cat\'{o}lica del Per\'{u}, Lima, Peru\\
$^{101}$ Stefan Meyer Institut f\"{u}r Subatomare Physik (SMI), Vienna, Austria\\
$^{102}$ SUBATECH, IMT Atlantique, Nantes Universit\'{e}, CNRS-IN2P3, Nantes, France\\
$^{103}$ Sungkyunkwan University, Suwon City, Republic of Korea\\
$^{104}$ Suranaree University of Technology, Nakhon Ratchasima, Thailand\\
$^{105}$ Technical University of Ko\v{s}ice, Ko\v{s}ice, Slovak Republic\\
$^{106}$ The Henryk Niewodniczanski Institute of Nuclear Physics, Polish Academy of Sciences, Cracow, Poland\\
$^{107}$ The University of Texas at Austin, Austin, Texas, United States\\
$^{108}$ Universidad Aut\'{o}noma de Sinaloa, Culiac\'{a}n, Mexico\\
$^{109}$ Universidade de S\~{a}o Paulo (USP), S\~{a}o Paulo, Brazil\\
$^{110}$ Universidade Estadual de Campinas (UNICAMP), Campinas, Brazil\\
$^{111}$ Universidade Federal do ABC, Santo Andre, Brazil\\
$^{112}$ Universitatea Nationala de Stiinta si Tehnologie Politehnica Bucuresti, Bucharest, Romania\\
$^{113}$ University of Derby, Derby, United Kingdom\\
$^{114}$ University of Houston, Houston, Texas, United States\\
$^{115}$ University of Jyv\"{a}skyl\"{a}, Jyv\"{a}skyl\"{a}, Finland\\
$^{116}$ University of Kansas, Lawrence, Kansas, United States\\
$^{117}$ University of Liverpool, Liverpool, United Kingdom\\
$^{118}$ University of Science and Technology of China, Hefei, China\\
$^{119}$ University of South-Eastern Norway, Kongsberg, Norway\\
$^{120}$ University of Tennessee, Knoxville, Tennessee, United States\\
$^{121}$ University of the Witwatersrand, Johannesburg, South Africa\\
$^{122}$ University of Tokyo, Tokyo, Japan\\
$^{123}$ University of Tsukuba, Tsukuba, Japan\\
$^{124}$ Universit\"{a}t M\"{u}nster, Institut f\"{u}r Kernphysik, M\"{u}nster, Germany\\
$^{125}$ Universit\'{e} Clermont Auvergne, CNRS/IN2P3, LPC, Clermont-Ferrand, France\\
$^{126}$ Universit\'{e} de Lyon, CNRS/IN2P3, Institut de Physique des 2 Infinis de Lyon, Lyon, France\\
$^{127}$ Universit\'{e} de Strasbourg, CNRS, IPHC UMR 7178, F-67000 Strasbourg, France, Strasbourg, France\\
$^{128}$ Universit\'{e} Paris-Saclay, Centre d'Etudes de Saclay (CEA), IRFU, D\'{e}partment de Physique Nucl\'{e}aire (DPhN), Saclay, France\\
$^{129}$ Universit\'{e}  Paris-Saclay, CNRS/IN2P3, IJCLab, Orsay, France\\
$^{130}$ Universit\`{a} degli Studi di Foggia, Foggia, Italy\\
$^{131}$ Universit\`{a} del Piemonte Orientale, Vercelli, Italy\\
$^{132}$ Universit\`{a} di Brescia, Brescia, Italy\\
$^{133}$ Variable Energy Cyclotron Centre, Homi Bhabha National Institute, Kolkata, India\\
$^{134}$ Warsaw University of Technology, Warsaw, Poland\\
$^{135}$ Wayne State University, Detroit, Michigan, United States\\
$^{136}$ Yale University, New Haven, Connecticut, United States\\
$^{137}$ Yildiz Technical University, Istanbul, Turkey\\
$^{138}$ Yonsei University, Seoul, Republic of Korea\\
$^{139}$ Affiliated with an institute formerly covered by a cooperation agreement with CERN\\
$^{140}$ Affiliated with an international laboratory covered by a cooperation agreement with CERN.\\

\end{flushleft} 
  
\end{document}